\documentstyle[12pt]{article}

\author{Yu.~M.~Zinoviev
       \thanks{E-mail address: ZINOVIEV@MX.IHEP.SU} \\
        {\it Institute for High Energy Physics} \\
        {\it Protvino, Moscow Region, 142284, Russia}}
\title{Massive Spin-2 Supermultiplets}
\date{}

\begin{document}

\maketitle

\begin{abstract}
In this paper we construct explicit Lagrangian formulation for the
massive spin-2 supermultiplets with $N=k$ supersymmetries
$k =1,2,3,4$. Such multiplets contain $2k$ particles with spin-3/2, so
there must exist $N=2k$ local supersymmetries in the full nonlinear
theories spontaneously broken so that only $N=k$ global
supersymmetries remain unbroken. In this paper we unhide these hidden
supersymmetries by using gauge invariant formulation for massive high
spin particles. Such formulation, operating with the right set of
physical degrees of freedom from the very beginning and having non-
singular massless limit, turns out to be very well suited for
construction of massive supermultiplets from the well known massless
ones. For all four cases considered we have managed to show that the
massless limit of the supertransformations for $N=k$ massive
supermultiplet could be uplifted to $N=2k$ supersymmetry. This, in
turn, allows one to investigate which extended supergravity models
such massive multiplets could arise from. Our results show a clear
connection of possible models with the five-dimensional extended
supergravities.
\end{abstract}

\newpage
\setcounter{page}{1}

\section*{Introduction}

The problem of constructing a consistent interacting theory for
massive spin-2 particles is an old but still unsolved one. Massive
particle has more degrees of freedom than massless one and as a result
even small graviton mass could give observables consequences. So an
interesting and important question is how these additional degrees of
freedom interact with matter. One of the possible ways in this
direction is the investigation of supermultiplets containing massive
spin-2 particle. Supersymmetry (especially extended) is a very
restrictive symmetry so even the structure of free theories could give
interesting and useful information. Surprisingly, there exist quite a
few results on this subject. In \cite{BL97,BL99} all states of the
first massive level of four-dimensional superstring with $N=1$
supersymmetry was considered and superfield formulation for
massive spin-2 $N=1$ supermultiplet appeared in \cite{BGLP02}.

Massive spin-2 supermultiplets with $N=k$ supersymmetry contains $2k$
spin-3/2 particles \cite{S87,ADFL02a}. As consistent description of
every spin-3/2 particle requires local supersymmetry there must exist
$N=2k$ local supersymmetries spontaneously broken so that only $N=k$
global supersymmetries remain unbroken. In this paper we consider
Lagrangian formulation for all massive spin-2 supermultiplets for
$k=1,2,3,4$ (multiplets with $k>4$ will contain particles with spin
greater than 2). We unhide these hidden extended supersymmetries by
using gauge invariant description of massive high spin particles
\cite{Zin94,KZ97}. Such description operating with right number of
physical degrees of freedom from the very beginning and having non-
singular massless limit turns out to be well suited for the
investigation of massive high spin particles and their possible
interactions. It could be easily generalized to the higher dimensions
as well as (Anti)de Sitter space (e.g. \cite{Z01}).

Here we use a straightforward generalization to the case of massive
supermultiplets. One start with appropriate set of massless
supermultiplets, then adds all possible low derivative mass terms to
the Lagrangian as well as additional terms to the fermionic
supertransformation laws (bosonic supertransformations do not contain
derivatives, so there are no corrections to them). The requirement the
whole Lagrangian to be invariant under the supertransformations fixes
all the unknown coefficients in the Lagrangian and
supertransformations. By saying appropriate set of massless
supermultiplets we mean not only their number and particle content.
The presence of vector fields leads to the possibility to make duality
transformations mixing different supermultiplets. The existence of
such dual versions of extended supergravities plays very important
role in the problem of spontaneous supersymmetry breaking. As we will
see for the construction of massive spin-2 supermultiplets it turns
out absolutely necessary to choose the correct mixing of vector
fields.

In all four cases considered we have managed to show that the massless
limit of supertransformations of massive $N=k$ supermultiplet could be
uplifted up to $N=2k$ supersymmetry. This allows one to investigate
which extended supergravities such massive supermultiplets could arise
from and we give explicit examples of such theories having the correct
structure of vector fields mixing as well as global symmetries related
with the scalar fields. All examples turns out to be the theories that
can be obtained by dimensional reduction from the corresponding five-
dimensional supergravities.

In the following four sections we carry on such a program for massive
spin-2 supermultiplets with $N=1,2,3,4$ correspondingly. Our
notations, conventions and some useful formulas are collected in the
Appendix.

\section{$N=1$}

For $N=1$ supersymmetry massive spin-2 supermultiplet contains
\cite{S87,ADFL02a} four massive fields $(2, 3/2,3/2,1)$. As is well
known, in the massless limit massive spin-2 particle breaks into the
massless ones with spins 2, 1 and 0, massive spin-3/2 particle ---
into the massless spin-3/2 and spin-1/2, at last, massive spin-1 ---
into massless spin-1 and spin-0 particles. It is easy to see that in
this limit one gets just four massless $N=1$ supermultiplets, namely,
gravity multiplet $h_{\mu\nu},\Psi_\mu$, spin-3/2 multiplet
$\Phi_\mu,A_\mu$, vector multiplet $B_\mu.\chi$ and chiral multiplet
$\lambda,\varphi,\pi$. Note that chiral multiplet contains scalar and
pseudoscalar fields, which in the massive case play the role of the
Goldstone ones, so one of the vector fields has to be axial-vector.

The most general $N=1$ linear global supertransformations leaving the
sum of free massless Lagrangians invariant have the form:
\begin{eqnarray}
\delta h_{\mu\nu} &=& i (\bar{\Psi}_{(\mu} \gamma_{\nu)} \eta) \qquad
\delta \Psi_\mu = - \sigma^{\alpha\beta} \partial_\alpha h_{\beta\mu}
\eta  \nonumber \\
\delta \Phi_\mu &=& - \frac{i}{2\sqrt{2}} (\sigma (cos\theta A - sin
\theta \gamma_5 B)) \gamma_\mu \eta \nonumber \\
\delta A_\mu &=& \sqrt{2} cos\theta (\bar{\Phi}_\mu \eta) + i sin
\theta (\bar{\chi} \gamma_\mu \eta) \nonumber \\
\delta B_\mu &=& \sqrt{2} sin\theta (\bar{\Phi}_\mu \gamma_5 \eta) + i
cos\theta (\bar{\chi} \gamma_\mu \gamma_5 \eta) \\
\delta \chi &=& - \frac{1}{2} (\sigma(sin\theta A + cos\theta \gamma_5
B)) \eta \nonumber \\
\delta \lambda &=& - i \hat{\partial} (\varphi + \gamma_5 \pi) \eta
\qquad \delta \varphi = (\bar{\lambda} \eta) \qquad \delta \pi = (
\bar{\lambda} \gamma_5 \eta) \nonumber
\end{eqnarray}

One of the very important points here is the possibility to have a
mixing between vector field from the spin-3/2 multiplet and axial-
vector one from the vector multiplet, which is a manifestation of the
general duality symmetry of supersymmetric theories. As we will see
later, the construction of massive supermultiplet turns out to be
possible for one concrete value of mixing angle $\theta$ only.

Another very essential point is the requirement that the model be
invariant under the whole $U(N)$ $R$-symmetry of the superalgebra. In
the $N=1$ case it is just axial $U(1)$-symmetry, the axial charges of
all fields being as follows:
\begin{center}
\begin{tabular}{|c|c|c|c|} \hline
field & $\eta$, $\Psi_\mu$, $\chi$ & $\Phi_\mu$, $\lambda$ &
$h_{\mu\nu}$, $A_\mu$, $B_\mu$, $\varphi$, $\pi$ \\ \hline
axial charge & + 1 & - 1 & 0 \\ \hline
\end{tabular}
\end{center}
Now let us add to the Lagrangian the most general gauge invariant mass
terms (see Appendix) compatible with the axial $U(1)$ invariance:
\begin{eqnarray}
\frac{1}{m} {\cal L}_1 &=& \sqrt{2} [h^{\mu\nu} \partial_\mu A_\nu - h
(\partial A)] - \sqrt{3} A^\mu \partial_\mu \varphi - B^\mu \partial_
\mu \pi - \nonumber \\
  && - \bar{\Phi}_\mu \sigma^{\mu\nu} \Psi_\nu + i \kappa_1
  (\bar{\Psi} \gamma) \chi + i \kappa_2 (\bar{\Phi} \gamma) \lambda +
  \kappa_3 \bar{\chi} \lambda \\
 \frac{1}{m^2} {\cal L}_2 &=& - \frac{1}{2} (h^{\mu\nu} h_{\mu\nu} -
 h^2) - \sqrt{\frac{3}{2}} h \varphi + \varphi^2 + \frac{1}{2}
 B_\mu{}^2 \nonumber
\end{eqnarray}
Note, in particular, that axial $U(1)$ invariance dictates the Dirac
mass term for the gravitini. To make complete Lagrangian invariant
under the supertransformations one has to add new terms to the
fermionic transformation laws:
\begin{eqnarray}
\frac{1}{m} \delta' \Psi_\mu  &=& (\alpha_1 A_\mu + \alpha_2
\sigma_{\mu\nu} A^\nu + \alpha_3 \gamma_5 B_\mu) \eta \nonumber \\
\frac{1}{m} \delta' \Phi_\mu  &=& (i \alpha_4 h_{\mu\nu} \gamma^\nu +
i \alpha_5 \varphi \gamma_\mu + i \alpha_6 \pi \gamma_5 \gamma_\mu)
\eta \\
\frac{1}{m} \delta' \chi  &=& (\beta_1 \varphi + \beta_2 \gamma_5 \pi)
\eta \qquad \frac{1}{m} \delta' \lambda  = (i \beta_3 \hat{A} + i
\beta_4 \hat{B} \gamma_5) \eta \nonumber
\end{eqnarray}
Simple calculations show that the invariance fixes the mixing angle
$sin\theta = \sqrt{3}/2$, $cos\theta = 1/2$, as well as all
coefficients $\alpha$ and $\beta$:
\begin{eqnarray*}
\kappa_1 &=& \kappa_2 = \sqrt{\frac{3}{2}} \qquad \kappa_3 = - 2 \\
\alpha_1 &=& \alpha_2 = - \frac{1}{\sqrt{2}} \qquad \alpha_3 = -
\sqrt{\frac{3}{2}} \qquad \alpha_4 = 1 \qquad \alpha_5 = -
\frac{3}{2\sqrt{6}} \qquad \alpha_6 = \frac{3}{2\sqrt{6}} \\
\beta_1 &=& - \frac{1}{2} \qquad \beta_2 = - \frac{3}{2} \qquad
\beta_3 = \sqrt{3} \qquad \beta_4 = 1
\end{eqnarray*}
Apart from the $N=1$ global supertransformations given above the
Lagrangian is invariant under two local (spontaneously broken)
supertransformations:
\begin{eqnarray*}
\delta \Psi_\mu &=& \partial \xi_1 \qquad \delta \Phi_\mu =
\frac{im}{2} \gamma_\mu \xi_1 \qquad \delta \chi = \sqrt{\frac{3}{2}}
m \xi_1 \\
\delta \Phi_\mu &=& \partial \xi_2 \qquad \delta \Psi_\mu =
\frac{im}{2} \gamma_\mu \xi_2 \qquad \delta \lambda =
\sqrt{\frac{3}{2}} m \xi_2
\end{eqnarray*}
One can use these transformations to bring the fermionic laws to more
simple and convenient form. For example, making field dependent
$\xi_1$ transformation with $\xi_1 = \sqrt{3/2} \gamma_5 \pi \eta$ we
obtain:
\begin{eqnarray}
\delta \Psi_\mu &=& - \sigma^{\alpha\beta} \partial_\alpha
h_{\beta\mu} \eta + \sqrt{\frac{3}{2}} \gamma_5 D_\mu \pi \eta -
\frac{m}{\sqrt{2}} \gamma_\mu \hat{A} \eta \nonumber \\
\delta \Phi_\mu &=& - \frac{i}{4\sqrt{2}} (\sigma(A - \sqrt{3}
\gamma_5 B)) \eta + im h_{\mu\nu} \gamma^\nu \eta -
\frac{3im}{2\sqrt{6}} \varphi \gamma_\mu \eta \nonumber \\
\delta \chi &=& - \frac{1}{4} (\sigma(\sqrt{3} A + \gamma_5 B)) \eta -
\frac{m}{2} \varphi \eta \\
\delta \lambda &=& - i \gamma^\mu (\partial_\mu \varphi + \gamma_5
D_\mu \pi) \eta + i m \sqrt{3} \hat{A} \eta \nonumber
\end{eqnarray}
where $D_\mu \pi = \partial_\mu \pi - m B_\mu$.

A few comments are in order.

Now the (pseudo)scalar field $\pi$ enters the Lagrangian and
supertransformation laws through the derivative $\partial_\mu \pi$
only. So one can construct a dual formulation of such model where this
field is replaced by the skew-symmetric tensor $C_{[\mu\nu]}$.
Analogous construction for the massive spin-3/2 supermultiplet was
considered in \cite{AB99}. It seems that such formulation could be
closer to the results of \cite{BGLP02}.

As usual in gauge invariant formulation of massive high spin fields
one can use local gauge transformations to exclude all the Goldstone
fields setting the gauge $A_\mu = \chi = \lambda = \varphi = \pi = 0$.
In such a gauge one deals with four physical massive fields
$h_{\mu\nu}$, $\Psi_\mu$, $\Phi_\mu$ and $B_\mu$ only, but the
supertransformation leaving such Lagrangian invariant would be the
combination of usual supertransformations given above and (higher
derivative) field dependent gauge transformations restoring the gauge.
We prefer to work with the complete gauge invariant formulation
because it has a non-singular massless limit and so it is very well
suited for the investigation of supergravity models which such massive
supermultiplets could arise from.

In the massless limit the set of fields used perfectly combines into
just two $N=2$ supermultiplets, namely $N=2$ supergravity multiplet
and one vector multiplet. Let us stress however that it does not mean
that $N=1$ vector and chiral multiplets should belong to the same
$N=2$ supermultiplet. Recall that very similar situation appears when
one consider the partial super-Higgs effect in $N=2$ supergravity. The
massless limit of $N=1$ supergravity plus massive $N=1$ spin-3/2
supermultiplets gives exactly the same set of fields. But to construct
the whole interacting theory one has to assign $N=1$ vector and chiral
multiplets to two different $N=2$ supermultiplets, namely to vector
and hypermultiplet \cite{CGP86,Z87}

Nevertheless, it is interesting to check the existence of minimal
model of $N=2$ supergravity with just one vector multiplet having
desired properties, the most important of which being the correct
mixing of vector and axial-vector fields and global symmetries related
with the scalar fields $\varphi$ and $\pi$. Now we will show that such
a model really exists. First of all we take massless limit of our
$N=1$ supertransformations and uplift them up to $N=2$
supertransformations:
\begin{eqnarray}
\delta e_{\mu a} &=& i (\bar{\Psi}_\mu{}^i \gamma_a \eta_i) \nonumber
\\
\delta \Psi_{\mu i} &=& 2 D_\mu \eta_i - \frac{i}{4\sqrt{2}}
\varepsilon_{ij} (\sigma(A - \sqrt{3} \gamma_5 B)) \gamma_\mu \eta^j -
\sqrt{\frac{3}{2}} \gamma_5 \partial_\mu \pi \eta_i \nonumber \\
\delta A_\mu &=& \frac{1}{\sqrt{2}} \varepsilon^{ij}
(\bar{\Psi}_{\mu i} \eta_j) + i \frac{\sqrt{3}}{2} (\bar{\chi}^i
\gamma_\mu \eta_i) \\
\delta B_\mu &=& \sqrt{\frac{3}{2}} \varepsilon^{ij}
(\bar{\Psi}_{\mu i} \gamma_5 \eta_j) + \frac{i}{2} (\bar{\chi}^i
\gamma_\mu \gamma_5 \eta_i) \nonumber \\
\delta \chi_i &=& - \frac{1}{4} (\sigma(\sqrt{3} A + \gamma_5 B))
\eta_i - i \varepsilon_{ij} \hat{\partial} (\varphi + \gamma_5 \pi)
\eta^j \nonumber \\
\delta \varphi &=& \varepsilon^{ij} (\bar{\chi}_i \eta_j) \qquad
\delta \pi = \varepsilon^{ij} (\bar{\chi}_i \gamma_5 \eta_j) \nonumber
\end{eqnarray}
where $\Psi_{\mu i} = (\Psi_\mu, \Phi_\mu)$,
$\chi_i = (\chi, \lambda)$. Now one can use straightforward Noether
procedure to obtain complete nonlinear interacting Lagrangian. The
bosonic part of this Lagrangian appears to be:
\begin{eqnarray}
{\cal L}_B &=& - \frac{1}{2} R + \frac{1}{2} \partial_\mu \varphi
 \partial_\mu \varphi + \frac{1}{2} \Phi^{-4} \partial_\mu \pi
 \partial_\mu \pi - \frac{1}{4} \Phi^6 A_{\mu\nu}{}^2 -
\frac{1}{4} \Phi^2 (B_{\mu\nu} - \sqrt{2} \pi A_{\mu\nu})^2 -
\nonumber \\
 && - \frac{1}{4\sqrt{6}} \pi [B_{\mu\nu} \tilde{B}_{\mu\nu} -
 \sqrt{2} \pi A_{\mu\nu} \tilde{B}_{\mu\nu} + \frac{2}{3} \pi^2
 A_{\mu\nu} \tilde{A}_{\mu\nu} ]
\end{eqnarray}
while bilinear in fermionic fields part looks as follows:
\begin{eqnarray}
{\cal L}_{2F} &=& \frac{i}{2} \varepsilon^{\mu\nu\alpha\beta}
\bar{\Psi}_\mu{}^i \gamma_5 \gamma_\nu D_\alpha \Psi_{\beta i} + \frac
{i}{2} \bar{\chi}^i \gamma^\mu D_\mu \chi_i - \nonumber \\
 && - \frac{1}{4\sqrt{2}} \varepsilon^{ij}
\bar{\Psi}_{\mu i} [ \Phi^3 (A^{\mu\nu} - \gamma_5 \tilde{A}^{\mu\nu})
+ \sqrt{3} \Phi (\gamma_5 B^{\mu\nu} + \tilde{B}^{\mu\nu}) - \sqrt{6}
\Phi \pi (\gamma_5 A^{\mu\nu} + \tilde{A}^{\mu\nu})] \Psi_{\nu j} +
\nonumber \\
 && + \frac{i}{8} \bar{\chi}^i \gamma^\mu (\sigma [\sqrt{3} \Phi^3 A +
\gamma_5 \Phi B - \sqrt{2} \gamma_5 \Phi \pi A]) \Psi_{\mu i} -
\nonumber \\
 && - \frac{1}{4\sqrt{6}} \varepsilon^{ij} \bar{\chi}_i (\sigma [
\sqrt{3} \Phi^3 A + \gamma_5 \Phi B - \sqrt{2} \gamma_5 \Phi \pi A])
\chi_j - \nonumber \\
 && - \frac{1}{2} \varepsilon^{ij} \bar{\chi}_i \gamma^\mu \gamma^
\nu (\partial_\nu \varphi + \gamma_5 \Phi^{-2} \partial_\nu \pi)
\Psi_{\mu j} - \nonumber \\
 && - \frac{i}{4} \Phi^{-2} [ \sqrt{\frac{3}{2}}
\varepsilon^{\mu\nu\alpha\beta} \bar{\Psi}_\mu{}^i \gamma_\nu
\Psi_{\beta i} + \frac{1}{\sqrt{6}} \bar{\chi}^i \gamma^\alpha
\gamma_5 \chi_i ] \partial_\alpha \pi
\end{eqnarray}
Here $\Phi = exp(- \frac{1}{\sqrt{6}} \varphi)$. This Lagrangian is
invariant under the following $N=2$ local supertransformations:
\begin{eqnarray}
\delta e_{\mu a} &=& i (\bar{\Psi}_\mu{}^i \gamma_a \eta_i) \nonumber
\\
\delta \Psi_{\mu i} &=& 2 D_\mu \eta_i - \frac{i}{4\sqrt{2}}
\varepsilon_{ij} (\sigma[\Phi^3 A - \sqrt{3} \gamma_5 \Phi (B - \sqrt{
2} \pi A)]) \gamma_\mu \eta^j - \sqrt{\frac{3}{2}} \gamma_5 \Phi^{-2}
\partial_\mu \pi \eta_i \nonumber \\
\delta A_\mu &=& \frac{1}{\sqrt{2}} \Phi^{-3} \varepsilon^{ij}
(\bar{\Psi}_{\mu i} \eta_j) + i \frac{\sqrt{3}}{2} \Phi^{-3}
(\bar{\chi}^i \gamma_\mu \eta_i) \\
\delta B_\mu &=& \sqrt{\frac{3}{2}} \Phi^{-1} \varepsilon^{ij}
(\bar{\Psi}_{\mu i} \gamma_5 \eta_j) + \frac{i}{2} \Phi^{-1}
(\bar{\chi}^i \gamma_\mu \gamma_5 \eta_i) + \sqrt{2} \pi \delta A_\mu
\nonumber \\
\delta \chi_i &=& - \frac{1}{4} (\sigma[\sqrt{3} \Phi^3 A + \gamma_5
(B - \sqrt{2} \pi A)]) - i \varepsilon_{ij} \gamma^\mu (\partial_\mu
\varphi + \gamma_5 \Phi^{-2} \partial_\mu \pi) \eta^j
\nonumber \\
\delta \varphi &=& \varepsilon^{ij} (\bar{\chi}_i \eta_j) \qquad
\delta \pi = \Phi^2 \varepsilon^{ij} (\bar{\chi}_i \gamma_5 \eta_j)
\nonumber
\end{eqnarray}

Apart from the local supertransformations this Lagrangian is invariant
under two global transformations. One of them is a translation:
\begin{equation}
\pi \rightarrow \pi + \tilde{\Lambda} \qquad B_\mu \rightarrow B_\mu +
\sqrt{2} A_\mu \tilde{\Lambda} \label{trans}
\end{equation}
while another one is a scale transformation:
\begin{equation}
\Phi \rightarrow e^\Lambda \Phi \qquad A_\mu \rightarrow e^{-3\Lambda}
A_\mu \qquad B_\mu \rightarrow e^{-\Lambda} B_\mu \qquad \pi
\rightarrow e^{2\Lambda} \pi \label{scale}
\end{equation}

In extended supergravities the only way to obtain nontrivial scalar
field potential, symmetry breaking, masses and so on is the gauging
of (part of) the global symmetries. The gauging of the translations
like (\ref{trans}) is now rather well known mechanism playing very
important role in the problem of spontaneous supersymmetry breaking
without a cosmological term
\cite{Z87,Z92,Z95,Z86,TZ96a,Z87a,TZ96b,ADFL02a,ADFL02b,ADFL02c}. But
the gauging of scale transformation like (\ref{scale}) in the usual
Poincare supergravities is an open question that requires further
investigations.

The model of $N=2$ supergravity with vector multiplet constructed here
is one of the simplest examples of the so called no-scale models
\cite{CKPDFWG85} tightly connected with the reduction of $N=2$ $D=5$
supergravities \cite{GST84}. Such connection with $D=5$ supergravities
could hardly be a surprise. In the same way as massive $D=4$ particles
can be constructed by the reduction from massless $D=5$ ones, massive
$D=4$ supermultiplets could arise from the reduction of appropriate
$D=5$ ones. For the massive spin-3/2 supermultiplet such procedure was
considered in \cite{A99}. The possibility to obtain full nonlinear
massive spin-2 $D=4$ theory from massless $D=5$ supergravity by some
kind of compactification (for example radial compactification a la
\cite{BS02}) is again an open question.

\section{$N=2$}

For $N=2$ massive spin-2 supermultiplet contains following particles
$(2, 4 \times 3/2, 6 \times 1, 4 \times 1/2, 0)$. In the massless
limit we get additionally $(1, 4 \times 1/2, 7 \times 0)$ Goldstone
particles. It is easy to see that in terms of $N=2$ massless
supermultiplets we have $N=2$ supergravity $(2, 2 \times 3/2, 1)$,
doublet of spin-3/2 supermultiplets $2 \times (3/2, 2 \times 1, 1/2)$,
two vector multiplets $2 \times (1, 2 \times 1/2, 2 \times 0)$ and one
hypermultiplet $(2 \times 1/2, 4 \times 0)$.

As is well known usual $N=2$ supergravity is invariant under the
$SU(2)$ subgroup of full $U(2)$ $R$-symmetry group only. But there
exist a dual version of the system $N=2$ supergravity with one vector
multiplet where graviphoton of supergravity multiplet and vector
field from the vector multiplet enter in the complex combination
providing $U(2)$ symmetry. So as our first building block we choose
the following mixed version of supergravity --- vector multiplet:
\begin{eqnarray}
\delta h_{\mu\nu} &=& i (\bar{\Psi}_{(\mu}{}^i \gamma_{\nu)} \eta_i)
\nonumber \\
\delta \Psi_{\mu i} &=& - \sigma^{\alpha\beta} \partial_\alpha
h_{\beta\mu} \eta_i - \frac{i}{4} (\sigma(C_4 - \gamma_5 C_5))
\varepsilon_{ij} \gamma_\mu \eta^j \nonumber \\
\delta C_{4\mu} &=& \varepsilon^{ij} (\bar{\Psi}_{\mu i} \eta_j) +
\frac{i}{\sqrt{2}} \varepsilon^{ij} (\bar{\lambda}^i \gamma_\mu
\eta_j) \\
\delta C_{5\mu} &=& \varepsilon^{ij} (\bar{\Psi}_{\mu i} \gamma_5
\eta_j) + \frac{i}{\sqrt{2}} \varepsilon^{ij} (\bar{\lambda}^i
\gamma_\mu \gamma_5 \eta_j) \nonumber \\
\delta \lambda^i &=& - \frac{1}{2\sqrt{2}} (\sigma(C_4 + \gamma_5
C_5)) \varepsilon^{ij} \eta_j - i \gamma^\mu \partial_\mu (\varphi +
\gamma_5 \pi) \eta_i \nonumber \\
\delta \varphi &=& (\bar{\lambda}^i \eta_i) \qquad \delta \pi =
(\bar{\lambda}^i \gamma_5 \eta_i) \nonumber
\end{eqnarray}

Now let us turn to the spin-3/2 supermultiplets. In order to have a
possibility to introduce $SU(2)$ invariant (Dirac) mass term for all
four gravitini the two spin-3/2 fields of this multiplets should
transformed as a doublet under $SU(2)$. In this, vector fields are
transformed as a triplet and a singlet. As in the $N=1$ case it is
crucial for the whole construction that one can introduce a mixing
between this singlet vector field and a vector field from the
remaining vector supermultiplet. Thus our second building block is the
doublet of spin-3/2 supermultiplet mixed with one vector
supermultiplet:
\begin{eqnarray}
\delta \Phi_\mu{}^i &=& - \frac{i}{4} (\sigma C)^a \gamma_\mu
(\tau^a)_i{}^j \eta_j - \frac{i}{4} (\sigma(sin\theta A - cos\theta
\gamma_5 B)) \gamma_\mu \eta_i \nonumber \\
\delta C_\mu{}^a &=& (\bar{\Phi}_\mu{}^i (\tau^a)_i{}^j \eta_j) +
\frac{i}{\sqrt{2}} (\bar{\lambda}_i \gamma_\mu (\tau^a)_i{}^j \eta_j)
\nonumber \\
\delta A_\mu &=& sin\theta (\bar{\Phi}_\mu{}^i \eta_i) +
\frac{i}{\sqrt{2}} sin\theta (\bar{\lambda}_i \gamma_\mu \eta_i) + i
cos\theta (\bar{\chi}_i \gamma_\mu \eta_i) \nonumber \\
\delta B_\mu &=& cos\theta (\bar{\Phi}_\mu{}^i \gamma_5 \eta_i) -
\frac{i}{\sqrt{2}} cos\theta (\bar{\lambda}_i \gamma_\mu \gamma_5
\eta_i) + i sin\theta (\bar{\chi}_i \gamma_\mu \gamma_5 \eta_i) \\
\delta \lambda_i &=& - \frac{1}{2\sqrt{2}} (\sigma C)^a (\tau^a)_i{}^j
\eta_j - \frac{1}{2\sqrt{2}} (\sigma(sin\theta A - cos\theta \gamma_5
B)) \eta_i \nonumber \\
\delta \chi_i &=& - \frac{1}{2} (\sigma(cos\theta A + sin\theta
\gamma_5 B)) \eta_i - i \varepsilon_{ij} \gamma^\mu \partial_\mu (z_4
+ \gamma_5 z_5) \eta^j \nonumber \\
\delta z_4 &=& \varepsilon^{ij} (\bar{\chi}_i \eta_j) \qquad \delta
z_5 = \varepsilon^{ij} (\bar{\chi}_i \gamma_5 \eta_j) \nonumber
\end{eqnarray}
here $a = 1,2,3$, and $(\tau^a)_i{}^j$ antihermitian $2 \times 2$
matrices, normalized so that $Sp (\tau^a \tau^b) = - 2 \delta^{ab}$.

The last block is just the hypermultiplet which we choose in the
formulation with doublet of spinor fields and triplet and singlet of
scalars (so called linear multiplet):
\begin{eqnarray}
\delta \chi^i &=& - i \gamma^\mu \partial_\mu (\tilde{\varphi}
\delta_i{}^j + z^a (\tau^a)_i{}^j ) \eta_j \nonumber \\
\delta \tilde{\varphi} &=& (\bar{\chi}^i \eta_i) \qquad \delta z^a = (
\bar{\chi}^i (\tau^a)_i{}^j \eta_j)
\end{eqnarray}

The structure of these supertransformations unambiguously fixes the
axial charges of all fields:
\begin{center}
\begin{tabular}{|c|c|c|c|c|} \hline
field & $\eta_i$, $\Psi_{\mu i}$, $\lambda_i$, $\chi_i$ &
$\Phi_\mu{}^i$, $\lambda^i$, $\chi^i$ & $C_4 + \gamma_5 C_5$,
$z_4 + \gamma_5 z_5$ & others \\ \hline
axial charge & + 1 & - 1 & - 2 & 0 \\ \hline
\end{tabular}
\end{center}

By using $SU(2)$ and $U(1)$ properties of all fields we can choose
which (combination of) fields will play the role of the Goldstone
ones. For the massive spin-2 $h_{\mu\nu}$ it has to be vector field
$A_\mu$ and some combination of two scalars $\varphi$ and
$\tilde{\varphi}$. For spin-3/2 field $\Psi_{\mu i}$ it could be
combination of spinors $\lambda_i$ and $\chi_i$, while for the
$\Phi_\mu{}^i$ --- combination of $\lambda^i$ and $\chi^i$. At last
for vector fields $C_\mu{}^a$, $C_{4\mu}$, $C_{5\mu}$, $B_\mu$ it will
be $z^a$, $z_{4,5}$ and $\pi$ correspondingly. Having made this
assignment we can write the most general mass terms compatible with
the $U(2)$ symmetry:
\begin{eqnarray}
\frac{1}{m} {\cal L}_1 &=& \sqrt{2} [h^{\mu\nu} \partial_\mu A_\nu - h
(\partial A)] - \sqrt{3} A^\mu \partial_\mu \varphi_1 - B_\mu
\partial_\mu \pi - C_\mu{}^a \partial_\mu z^a - C_4{}^\mu \partial_\mu
z_4 - C_5{}^\mu \partial_\mu z_5 - \nonumber \\
 && - \bar{\Phi}_\mu{}^i \sigma^{\mu\nu} \Psi_{\nu i} + i \kappa_1
 \bar{\Psi}_\mu{}^i \gamma^\mu \chi_i + i \kappa_2 \bar{\Phi}_{\mu i}
 \gamma^\mu \lambda^i + i \kappa_5 \bar{\Psi}_\mu{}^i \gamma^\mu
 \lambda_i + i \kappa_6 \bar{\Phi}_{\mu i} \gamma^\mu \chi^i +
 \nonumber \\
 && + \kappa_3 \bar{\lambda}^i \chi_i + \kappa_4 \bar{\chi}^i
 \lambda_i + \kappa_7 \bar{\chi}^i \chi_i + \kappa_8 \bar{\lambda}^i
 \lambda_i \\
 \frac{1}{m^2} {\cal L}_2 &=& - \frac{1}{2} (h^{\mu\nu} h_{\mu\nu} -
 h^2) - \sqrt{\frac{3}{2}} h \varphi_1 + \varphi_1{}^2 + \frac{1}{2}
 B_\mu{}^2 + \nonumber \\
 &&  + \frac{1}{2} (C_\mu{}^a)^2 +\frac{1}{2} (C_{4\mu})^2 +
 \frac{1}{2} (C_{5\mu})^2 - \frac{1}{2} \varphi_2{}^2
\end{eqnarray}
where $\varphi_1 = cos\alpha \varphi + sin\alpha \tilde{\varphi}$,
$\varphi_2 = - sin\alpha \varphi + cos\alpha \tilde{\varphi}$. To
restore the invariance under the global $N=2$ supertransformations we
have add appropriate terms to the fermionic transformation laws (see
later). In this, supersymmetry fixes both mixing angles:
$cos\theta = sin\theta = 1/\sqrt{2}$, $cos\alpha = 1/\sqrt{3}$,
$sin\alpha = \sqrt{2/3}$, all the coefficients for the mass terms in
the Lagrangian:
$$
\kappa_1 = 1, \qquad \kappa_2 = \kappa_5 = \frac{1}{\sqrt{2}}, \qquad
\kappa_3 = \kappa_4 = - \sqrt{2}, \qquad \kappa_6 = - \kappa_7 = 1,
\qquad \kappa_8 = 0
$$
as well as all the coefficients in the fermionic transformation laws.
The structure of the fermionic mass terms corresponds to two pairs of
local gauge symmetries:
\begin{eqnarray*}
\delta \Psi_{\mu i} &=& \partial_\mu \eta_i \qquad \delta \Phi_\mu{}^i
= \frac{im}{2} \gamma_\mu \eta_i \qquad \delta \lambda_i =
\frac{m}{\sqrt{2}} \eta_i \qquad \delta \chi_i = m \eta_i \\
\delta \Psi_{\mu i} &=& \frac{im}{2} \gamma_\mu \xi^i \qquad \delta
\Phi_\mu{}^i = \partial_\mu \xi^i  \qquad \delta \lambda^i =
\frac{m}{\sqrt{2}} \xi^i \qquad \delta \chi^i = m \xi^i \\
\end{eqnarray*}
One can use these freedom to make (field dependent) local
transformations to bring supertransformations of fermions to most
simple and convenient form:
\begin{eqnarray}
\delta \Psi_{\mu i} &=& - \sigma^{\alpha\beta} \partial_\alpha
h_{\beta\mu} \eta_i - \frac{i}{4} (\sigma(C_4 - \gamma_5 C_5))
\varepsilon_{ij} \gamma_\mu \eta^j + D_\mu z^a (\tau^a)_i{}^j \eta_j +
\frac{1}{\sqrt{2}} \gamma_5 D_\mu \pi \eta_i - \frac{m}{\sqrt{2}}
\gamma_\mu \hat{A} \eta_i \nonumber \\
\delta \Phi_\mu{}^i &=& - \frac{i}{4} (\sigma C)^a \gamma_\mu
(\tau^a)_i{}^j \eta_j - \frac{i}{4\sqrt{2}} (\sigma(A - \gamma_5 B))
\gamma_\mu \eta_i + D_\mu (z_4 + \gamma_5 z_5) \varepsilon^{ij} \eta_j
+ \nonumber \\
 && + i m h_{\mu\nu} \gamma^\nu \eta_i - \frac{im}{2} \gamma_\mu
 ( \frac{1}{\sqrt{2}} \varphi + \tilde{\varphi}) \eta_i \nonumber \\
\delta \lambda_i &=& - \frac{1}{2\sqrt{2}} (\sigma C)^a (\tau^a)_i{}^j
\eta_j - \frac{1}{4} (\sigma(A - \gamma_5 B)) \eta_i + \frac{m}{2}
\varphi \eta_i - \frac{m}{\sqrt{2}} \tilde{\varphi} \eta_i \\
\delta \chi_i &=& - \frac{1}{2\sqrt{2}} (\sigma(A + \gamma_5 B))
\eta_i - i \varepsilon_{ij} \gamma^\mu D_\mu (z_4 + \gamma_5 z_5)
\eta^j - \frac{m}{\sqrt{2}} \varphi \eta_i \nonumber \\
\delta \lambda^i &=& - \frac{1}{2\sqrt{2}} (\sigma(C_4 + \gamma_5
C_5)) \varepsilon^{ij} \eta_j - i \gamma^\mu (\partial_\mu \varphi +
\gamma_5 D_\mu \pi) \eta_i + i m \hat{A} \eta_i \nonumber \\
\delta \chi^i &=& - i \gamma^\mu (\partial_\mu \tilde{\varphi}
\delta_i{}^j + D_\mu z^a (\tau^a)_i{}^j ) \eta_j + i m \sqrt{2}
\hat{A} \eta_i \nonumber
\end{eqnarray}
where $D_\mu z^a = \partial_\mu z^a - m C_\mu{}^a$ and so on.

Now let us turn to the question which $N=4$ supergravity theory such a
massive supermultiplet could originate from. As we already mentioned
at the beginning of this section in the massless limit we have the
following massless $N=2$ supermultiplets: $N=2$ supergravity, doublet
of spin-3/2 supermultiplets, two vector and one hypermultiplet.
Certainly, to have $N=4$ supergravity one has to combine $N=2$
supergravity, two spin-3/2 and one vector supermultiplets. This leaves
us with one vector and one hypermultiplet. Not in any way evident that
they have to come from one and the same $N=4$ vector supermultiplet,
but it is the simplest possibility, so we restrict ourselves to the
models of $N=4$ supergravity with one vector supermultiplet. As is
well know due to $O(6) \approx SU(4)$ there exist $SU(4)$-invariant
formulation of $N=4$ supergravity \cite{CSF78} . Moreover, it could
consistently couples to the arbitrary number of vector
supermultiplets, the scalar field geometry being
$SO(6,n)/SO(6) \otimes SO(n)$. But the most general coupling of $N=4$
supergravity with vector supermultiplets constructed in \cite{TZ96c}
heavily depends on the possibility to have a dual mixing of vector
fields from the matter supermultiplets with the graviphotons. In this,
though the scalar fields geometry remains to be the same, the global
symmetry of the whole Lagrangian is in general lower than $SO(6,n)$.
As we have seen above the construction of massive spin-2
supermultiplet requires the mixing of matter vector field with one of
the graviphotons. So as a maximum we could have $SO(5) \approx USp(4)$
symmetry. Now we will show that such a formulation of $N=4$
supergravity with one vector supermultiplet does exist.

First of all, we combine fermions to quartets:
$\Psi_{\mu \hat{i}} = (\Psi_{\mu i}, \Phi_\mu{}^i)$,
$\chi_{\hat{i}} = (\chi_i,\chi^i)$,
$\lambda^{\hat{i}} = (\lambda^i, \lambda_i)$, where
$\hat{i} = 1,2,3,4$. Then we introduce quintet of vector fields
$C_\mu{}^{\hat{a}} = (C_\mu{}^a,C_\mu{}^4,C_\mu{}^5)$,
$\hat{a} = 1,2,3,4,5$, as well as quintet of scalars
$z^{\hat{a}} = (z^a, z_4,z_5)$. At last we construct five
skew symmetric $4 \times 4$ matrices
$(\tau^{\hat{a}})_{\hat{i}\hat{j}}$:
$$
\tau_{1,2,3} = \left( \begin{array}{cc} 0 & \tau_{1,2,3} \\ -
\tau_{1,2,3} & 0 \end{array} \right) \qquad \tau_4 = \left(
\begin{array}{cc} \varepsilon_{ij} & 0 \\ 0 & - \varepsilon^{ij}
\end{array} \right) \qquad \tau_5 = \left( \begin{array}{cc} -
\varepsilon_{ij} \gamma_5 & 0 \\ 0 & - \varepsilon^{ij} \gamma_5
\end{array} \right)
$$
Now the massless limit of supertransformations obtained above could be
uplifted to $N=4$ (omitting hats):
\begin{eqnarray}
\delta e_{\mu r} &=& i (\bar{\Psi}_\mu{}^i \gamma_r \eta_i) \nonumber
\\
\delta \Psi_{\mu i} &=& 2 D_\mu \eta_i - \frac{i}{4} (\sigma C)^a
\gamma_\mu (\bar{\tau}^a)^{ij} \eta_j - \frac{i}{4\sqrt{2}} (\sigma(A
- \gamma_5 B)) \gamma_\mu \Omega^{ij} \eta_j + \nonumber \\
 && + \partial_\mu z^a \Omega_{ij} (\bar{\tau}^a)^{jk} \eta_k +
 \frac{1}{\sqrt{2}} \gamma_5 \partial_\mu \pi \eta_i \nonumber \\
\delta C_\mu{}^a &=& (\bar{\Psi}_{\mu i} (\bar{\tau}^a)^{ij} \eta_j) +
\frac{i}{\sqrt{2}} (\bar{\lambda}_i \gamma_\mu (\bar{\tau}^a)^{ij}
\eta_j) \nonumber \\
\delta A_\mu &=& \frac{1}{\sqrt{2}} (\bar{\Psi}_{\mu i} \Omega^{ij}
\eta_j) + \frac{i}{2} (\bar{\lambda}_i \gamma_\mu \Omega^{ij} \eta_j)
+ \frac{i}{\sqrt{2}} (\bar{\chi}^i \gamma_\mu \eta_i) \nonumber \\
\delta B_\mu &=& \frac{1}{\sqrt{2}} (\bar{\Psi}_{\mu i} \gamma_5
\Omega^{ij} \eta_j) - \frac{i}{2} (\bar{\lambda}_i \gamma_\mu \gamma_5
\Omega^{ij} \eta_j) + \frac{i}{\sqrt{2}} (\bar{\chi}^i \gamma_\mu
\gamma_5 \eta_i)  \\
\delta \lambda^i &=& - \frac{1}{2\sqrt{2}} (\sigma C)^a
(\bar{\tau}^a)^{ij} \eta_j - \frac{1}{4} (\sigma(A - \gamma_5 B))
\Omega^{ij} \eta_j - i \gamma^\mu \partial_\mu (\varphi + \gamma_5
\pi) \eta_i \nonumber \\
\delta \chi_i &=& - \frac{1}{2\sqrt{2}} (\sigma(A + \gamma_5 B))
\eta_i - i \gamma^\mu \partial_\mu (\tilde{\varphi} \Omega^{ij} + z^a
(\bar{\tau}^a)^{ij} ) \eta_j \nonumber \\
\delta \varphi &=& (\bar{\lambda}^i \eta_i) \qquad \delta \pi =
(\bar{\lambda}^i \gamma_5 \eta_i) \qquad \delta \tilde{\varphi} =
\Omega^{ij} (\bar{\chi}_i \eta_j) \qquad \delta z^a = (\bar{\chi}_i
(\bar{\tau}^a)^{ij} \eta_j) \nonumber
\end{eqnarray}
Here $\Omega^{[ij]}$ is skew symmetric $USp(4)$ invariant tensor.

Now it is straightforward task by using standard Noether procedure to
construct full nonlinear interacting theory. The bosonic part of the
Lagrangian turns out to be:
\begin{eqnarray}
{\cal L}_B &=& - \frac{1}{2} R + \frac{1}{2} (\partial_\mu \varphi)^2
+ \frac{1}{2} \Phi^{-4} (\partial_\mu \pi)^2 + \frac{1}{2}
(\partial_\mu \tilde{\varphi})^2 + \frac{1}{2} \tilde{\Phi}^{-2}
(\partial_\mu z^a)^2 - \nonumber \\
 && - \frac{1}{4} \tilde{\Phi}^2 \Phi^2 A_{\mu\nu}{}^2 -
\frac{1}{4} \tilde{\Phi}^2 \Phi^{-2} (B_{\mu\nu} + \sqrt{2} \pi A_{\mu
\nu})^2 - \frac{1}{4} \Phi^2 (C_{\mu\nu}{}^a - \sqrt{2} z^a
A_{\mu\nu})^2 + \nonumber \\
 && + \frac{\pi}{2\sqrt{2}} (C_{\mu\nu}{}^a - \sqrt{2} z^a
 A_{\mu\nu})(\tilde{C}_{\mu\nu}{}^a - \sqrt{2} z^a \tilde{A}_{\mu\nu})
 - \frac{z^a}{\sqrt{2}} C_{\mu\nu}{}^a \tilde{B}_{\mu\nu} +
 \frac{(z^a)^2}{2} A_{\mu\nu} \tilde{B}_{\mu\nu}
\end{eqnarray}
where $\Phi = exp(\frac{1}{\sqrt{2}} \varphi)$,
$\tilde{\Phi} = exp(-\tilde{\varphi})$. In this, bilinear in fermionic
fields part of the Lagrangian looks as follows:
\begin{eqnarray}
{\cal L}_{2F} &=& \frac{1}{4} \bar{\Psi}_{\mu i} (V^{\mu\nu} -
\gamma_5 \tilde{V}^{\mu\nu})^a (\bar{\tau}^a)^{ij} \Psi_{\nu j} -
\frac{1}{4\sqrt{2}} \bar{\Psi}_{\mu i} (U_{\mu\nu} - \gamma_5
\tilde{U}_{\mu\nu})^+ \Omega^{ij} \Psi_{\nu j} + \nonumber \\
 && + \frac{i}{4\sqrt{2}} \bar{\lambda}_i \gamma^\mu (\sigma V)^a (
 \bar{\tau}^a)^{ij} \Psi_{\mu j} + \frac{1}{8} \bar{\lambda}_i \gamma^
 \mu (\sigma U)^- \Omega^{ij} \Psi_{\mu j} + \nonumber \\
 && + \frac{i}{4\sqrt{2}} \bar{\chi}^i \gamma^\mu (\sigma U)^+ \Psi_{
 \mu i} + \frac{1}{4} \bar{\lambda}^i (\sigma U)^+ \chi_i
 - \frac{1}{8} \bar{\chi}_i [(\sigma V)^a (\bar{\tau}^a)^{ij} +
 \frac{1}{\sqrt{2}} (\sigma U)^- \Omega^{ij} ] \chi_j - \nonumber \\
 && - \frac{1}{2} \bar{\lambda}^i \gamma^\mu \gamma^\nu (\partial_\nu
 \varphi + \gamma_5 \Phi^{-2} \partial_\nu \pi) \Psi_{\mu i} -
 \frac{1}{2} \bar{\chi}_i \gamma^\mu \gamma^\nu (\partial_\nu
 \tilde{\varphi} \Omega^{ij} + \tilde{\Phi}^{-1} \partial_\nu z^a
 (\bar{\tau}^a)^{ij} ) \Psi_{\mu j} + \nonumber \\
 && + \frac{i}{4} \tilde{\Phi}^{-1} [ \varepsilon^{\mu\nu\alpha\beta}
 \bar{\Psi}_\mu{}^i \gamma_5 \gamma_\nu \Omega_{ij}
 (\bar{\tau}^a)^{jk}  \Psi_{\beta k} - \bar{\lambda}_i \gamma^\alpha (
 \bar{\tau}^a)^{ij}  \Omega_{jk} \lambda^k + \bar{\chi}^i \gamma^
 \alpha \Omega_{ij}  (\bar{\tau}^a)^{jk} \chi_k ] \partial_\alpha z^a
 + \nonumber \\
 && + \frac{i}{4\sqrt{2}} \Phi^{-2} [ \varepsilon^{\mu\nu\alpha\beta}
 \bar{\Psi}_\mu{}^i \gamma_\nu \Psi_{\beta i} + 3 \bar{\lambda}_i
 \gamma^\alpha \gamma_5 \lambda^i - \bar{\chi}^i \gamma^\alpha
 \gamma_5 \chi_i ] \partial_\alpha \pi
\end{eqnarray}
where we introduced the following notations:
\begin{eqnarray*}
U_{\mu\nu}{}^\pm &=& \tilde{\Phi} [ \Phi A_{\mu\nu} \pm \gamma_5
\Phi^{-1} (B_{\mu\nu} + \sqrt{2} \pi A_{\mu\nu})] \\
V_{\mu\nu}{}^a &=& \Phi (C_{\mu\nu}{}^a - \sqrt{2} z^a A_{\mu\nu})
\end{eqnarray*}

This Lagrangian is invariant under the following $N=4$ local
supertransformations:
\begin{eqnarray}
\delta e_{\mu r} &=& i (\bar{\Psi}_\mu{}^i \gamma_r \eta_i) \nonumber
\\
\delta \Psi_{\mu i} &=& 2 D_\mu \eta_i - \frac{i}{4} (\sigma V)^a
\gamma_\mu (\bar{\tau}^a)^{ij} \eta_j - \frac{i}{4\sqrt{2}}
(\sigma U)^- \gamma_\mu \Omega^{ij} \eta_j + \nonumber \\
 && + \tilde{\Phi}^{-1} \partial_\mu z^a \Omega_{ij} (\bar{\tau}^a)^{
 jk} \eta_k + \frac{1}{\sqrt{2}} \Phi^{-2} \gamma_5 \partial_\mu \pi
 \eta_i \nonumber \\
\delta C_\mu{}^a &=& \Phi^{-1} [ (\bar{\Psi}_{\mu i}
(\bar{\tau}^a)^{ij} \eta_j) + \frac{i}{\sqrt{2}} (\bar{\lambda}_i
\gamma_\mu (\bar{\tau}^a)^{ij} \eta_j) ] + \sqrt{2} z^a \delta A_\mu
\nonumber \\
\delta A_\mu &=& \Phi^{-1} \tilde{\Phi}^{-1}[ \frac{1}{\sqrt{2}}
(\bar{\Psi}_{\mu i} \Omega^{ij} \eta_j) + \frac{i}{2} (\bar{\lambda}_i
\gamma_\mu \Omega^{ij} \eta_j) + \frac{i}{\sqrt{2}} (\bar{\chi}^i
\gamma_\mu \eta_i) ] \nonumber \\
\delta B_\mu &=& \Phi \tilde{\Phi}^{-1} [ \frac{1}{\sqrt{2}}
(\bar{\Psi}_{\mu i} \gamma_5 \Omega^{ij} \eta_j) - \frac{i}{2}
(\bar{\lambda}_i \gamma_\mu \gamma_5 \Omega^{ij} \eta_j) +
\frac{i}{\sqrt{2}} (\bar{\chi}^i \gamma_\mu \gamma_5 \eta_i) ] - \sqrt
{2} \pi \delta A_\mu \\
\delta \lambda^i &=& \frac{1}{4} (\sigma U)^- \Omega^{ij} \eta_j -
\frac{1}{2\sqrt{2}} (\sigma V)^a (\bar{\tau}^a)^{ij} \eta_j - i
\gamma^\mu (\partial_\mu \varphi + \gamma_5 \Phi^{-2} \partial_\mu
\pi) \eta_i \nonumber \\
\delta \chi_i &=& - \frac{1}{2\sqrt{2}} (\sigma U)^+ \eta_i - i
\gamma_\mu (\partial_\mu \tilde{\varphi} \Omega^{ij} +
\tilde{\Phi}^{-1} \partial_\mu z^a (\bar{\tau}^a)^{ij} ) \eta_j
\nonumber \\
\delta \varphi &=& (\bar{\lambda}^i \eta_i) \qquad \delta \pi =
\Phi^2 (\bar{\lambda}^i \gamma_5 \eta_i) \qquad \delta \tilde{\varphi}
= \Omega^{ij} (\bar{\chi}_i \eta_j) \qquad \delta z^a = \tilde{\Phi}
(\bar{\chi}_i (\bar{\tau}^a)^{ij} \eta_j) \nonumber
\end{eqnarray}

The scalar fields geometry of this model is rather peculiar. Out of
eight scalar fields six are related with global translations:
$$
z^a \rightarrow z^a + \Lambda^a \qquad C_\mu{}^a \rightarrow C_\mu{}^a
+ \sqrt{2} A_\mu \Lambda^a \qquad \pi \rightarrow \pi + \Lambda_0
\qquad B_\mu \rightarrow B_\mu - \sqrt{2} A_\mu \Lambda_0
$$
while the remaining two correspond to two "scale" transformations:
$$
\Phi \rightarrow e^{\Lambda} \Phi \qquad A_\mu \rightarrow
e^{-\Lambda} A_\mu \qquad B_\mu \rightarrow e^\Lambda B_\mu \qquad C_
\mu{}^a \rightarrow e^{-\Lambda} C_\mu{}^a \qquad \pi \rightarrow
e^{2\Lambda} \pi
$$
$$
\tilde{\Phi} \rightarrow e^{\tilde{\Lambda}} \tilde{\Phi} \qquad A_\mu
\rightarrow e^{-\tilde{\Lambda}} A_\mu \qquad B_\mu \rightarrow
e^{-\tilde{\Lambda}} B_\mu \qquad z^a \rightarrow e^{\tilde{\Lambda}}
z^a
$$
As in the previous case, global symmetries of the Lagrangian,
including $USp(4)$ one, clearly shows that such a model is tightly
connected with the $N=4$ $D=5$ supergravity \cite{AT85,ADFL02c}.

\section{$N=3$}

In this case massive spin-2 supermultiplet contains the following
states: $(2, 6 \times 3/2, 15 \times 1, 20 \times 1/2, 14 \times 0)$.
In the massless limit one get additionally
$(1, 6 \times 1/2, 16 \times 0)$. Altogether we have
$(2, 6 \times 3/2, 16 \times 1, 26 \times 1/2, 30 \times 0)$ that
corresponds exactly to the spectrum of $N=6$ supergravity. But to
construct massive $N=3$ supermultiplet we have to start with $N=3$
massless supermultiplets. It is easy to see that we have $N=3$
supergravity multiplet $(2, 3 \times 3/2, 3 \times 1, 1/2)$, triplet
of spin-3/2 supermultiplet
$3 \times (3/2, 3 \times 1, 3 \times 1/2, 2 \times 0)$ and four vector
supermultiplets $4 \times (1, 4 \times 1/2, 6 \times 0)$.

As is well known usual $N=3$ supergravity is invariant under the real
$SO(3)$-subgroup of the whole $U(3)$ $R$-symmetry group only. But as
has been shown long time ago \cite{Z86} there exist dual version of
$N=3$ supergravity with three vector supermultiplets invariant under
the whole $U(3)$ group. It turned out that such a version admit the
spontaneous supersymmetry breaking with three arbitrary scales,
including partial super-Higgs effects $N=3 \rightarrow N=2$ and
$N=3 \rightarrow N=1$. Moreover this theory can be coupled to
arbitrary number of vector supermultiplets \cite{TZ96a}. So as our
first building block we choose right this system of $N=3$ supergravity
multiplet mixed with three vector supermultiplets:
\begin{eqnarray}
\delta h_{\mu\nu} &=& i (\bar{\Psi}_{(\mu}{}^i \gamma_{\nu)} \eta_i)
\nonumber \\
\delta \Psi_{\mu i} &=& - \sigma^{\alpha\beta} \partial_\alpha
h_{\beta\mu} \eta_i + \frac{i}{4} \varepsilon_{ijk} (\sigma(A -
\gamma_5 B))^j \gamma_\mu \eta^k \nonumber \\
\delta A_\mu{}^j &=& - \varepsilon^{ijk} (\bar{\Psi}_{\mu i} \eta_k) +
\frac{i}{\sqrt{2}} (\bar{\chi}^{ij} \gamma_\mu \eta_i) +
\frac{i}{\sqrt{2}} (\bar{\rho} \gamma_\mu \eta_j) \nonumber \\
\delta B_\mu{}^j &=& - \varepsilon^{ijk} (\bar{\Psi}_{\mu i} \gamma_5
\eta_k) + \frac{i}{\sqrt{2}} (\bar{\chi}^{ij} \gamma_\mu \gamma_5
\eta_i) - \frac{i}{\sqrt{2}} (\bar{\rho} \gamma_\mu \gamma_5 \eta_j)
\nonumber \\
\delta \chi_{ij} &=& - \frac{1}{2\sqrt{2}} (\sigma(A + \gamma_5 B))_j
\eta_i - \frac{i}{\sqrt{3}} \varepsilon_{ijk} \gamma^\mu \partial_\mu
(\varphi + \gamma_5 \pi) \eta^k - \frac{i}{\sqrt{2}} \varepsilon_{ikl}
\gamma^\mu \partial_\mu (\varphi_k{}^j + \gamma_5 \pi_k{}^j) \eta^l
\nonumber \\
\delta \rho &=& - \frac{1}{2\sqrt{2}} (\sigma(A - \gamma_5 B))^i
\eta_i \\
\delta \lambda^i &=& - \frac{i}{\sqrt{3}} \gamma^\mu \partial_\mu
(\varphi - \gamma_5 \pi) \eta_i - \frac{i}{\sqrt{2}} \gamma^\mu
\partial_\mu (\varphi_i{}^j - \gamma_5 \pi_i{}^j) \eta_j \nonumber \\
\delta \varphi &=& \frac{1}{\sqrt{3}} (\bar{\lambda}^i \eta_i) +
\frac{1}{\sqrt{3}} \varepsilon^{ijk} (\bar{\chi}_{ij} \eta_k) \qquad
\delta \pi = - \frac{1}{\sqrt{3}} (\bar{\lambda}^i \gamma_5 \eta_i) +
\frac{1}{\sqrt{3}} \varepsilon^{ijk} (\bar{\chi}_{ij} \gamma_5 \eta_k)
\nonumber \\
\delta \varphi^a &=& \frac{1}{\sqrt{2}} (\bar{\lambda}^i
(\tau^a)_i{}^j \eta_j) + \frac{1}{\sqrt{2}} \varepsilon^{ikl}
(\bar{\chi}_{ij} (\tau^a)_k{}^j \eta_l) \nonumber \\
\delta \pi^a &=& - \frac{1}{\sqrt{2}} (\bar{\lambda}^i \gamma_5
(\tau^a)_i{}^j \eta_j) + \frac{1}{\sqrt{2}} \varepsilon^{ikl}
(\bar{\chi}_{ij} \gamma_5 (\tau^a)_k{}^j \eta_l) \nonumber
\end{eqnarray}
here $\lambda^a$, $a = 1,2,...8$ --- eight hermitian $3 \times 3$
matrices normalized so that $Sp(\lambda^a \lambda^b) = 2 \delta^{ab}$.
In this, axial charges of all fields are:
\begin{center}
\begin{tabular}{|c|c|c|c|c|c|} \hline
field & $\Psi_{\mu i}$, $\eta_i$ & $\chi_{ij}$, $\lambda^i$ &
$(A_\mu + \gamma_5 B_\mu)_i$ & $\rho$ & others \\ \hline
axial charge & + 1 & - 1 & - 2 & + 3 & 0 \\ \hline
\end{tabular}
\end{center}

Now let us turn to the spin-3/2 multiplets. In order to have a
possibility to introduce $U(3)$ invariant (Dirac) mass terms for all
six gravitini the spin-3/2 fields of these multiplets should
transformed as antitriplet $\Phi_\mu{}^i$ under $SU(3)$. But then
vector fields are transformed as an octet $C_\mu{}^a$ and a singlet
$B_\mu$ leaving us with the possibility to make dual mixing of this
singlet vector with the vector fields of the remaining vector
supermultiplet. So our second (and the last) building block will be
(anti)triplet of spin-3/2 supermultiplet mixed with one vector
supermultiplet:
\begin{eqnarray}
\delta \Phi_\mu{}^i &=& \frac{i}{4} \gamma_5 (\sigma C)_j{}^i
\gamma_\mu \eta^j - \frac{i}{2\sqrt{6}} (\sigma (sin\theta A -
\gamma_5 cos\theta B)) \gamma_\mu \eta_i \nonumber \\
\delta C_\mu{}^a &=& (\bar{\Phi}_\mu{}^i \gamma_5 (\tau^a)_i{}^j
\eta_j) + \frac{i}{\sqrt{2}} (\bar{\chi}_{ij} \gamma_\mu \gamma_5
\varepsilon^{jkl} (\tau^a)_k{}^i \eta_l) \nonumber \\
\delta A_\mu &=& \frac{2}{\sqrt{6}} sin\theta (\bar{\Phi}_\mu{}^i
\eta_i) + \frac{i}{\sqrt{3}} sin\theta (\bar{\chi}_{ij} \gamma_\mu
\varepsilon^{ijk} \eta_k) + i cos\theta (\bar{\lambda}_i \gamma_\mu
\eta_i) \nonumber \\
\delta B_\mu &=& \frac{2}{\sqrt{6}} cos\theta (\bar{\Phi}_\mu{}^i
\gamma_5 \eta_i) - \frac{i}{\sqrt{3}} cos\theta (\bar{\chi}_{ij}
\gamma_\mu \gamma_5 \varepsilon^{ijk} \eta_k) + i sin\theta
(\bar{\lambda}_i \gamma_\mu \gamma_5 \eta_i) \nonumber \\
\delta \lambda_i &=& - \frac{1}{2} (\sigma (cos\theta A + \gamma_5
sin\theta B)) \eta_i - i \gamma^\mu \partial_\mu \varepsilon^{ijk}
\tilde{z}_j \eta_k \\
\delta \chi^{ij} &=& - \frac{1}{2\sqrt{2}} \varepsilon^{jkl} \gamma_5
(\sigma C)_k{}^i \eta_l - \frac{1}{2\sqrt{3}} (\sigma(sin\theta A -
\gamma_5 cos\theta B)) \varepsilon^{ijk} \eta_k - i \gamma^\mu
\partial_\mu z_i \eta_j \nonumber \\
\delta \tilde{\rho} &=& - i \gamma^\mu \partial_\mu \tilde{z}^{*i}
\eta_i \qquad \delta \varphi^i = (\bar{\chi}^{ij} \eta_j) \qquad
\delta \pi^i = (\bar{\chi}^{ij} \gamma_5 \eta_j) \nonumber \\
\delta \tilde{\varphi}^j &=& (\bar{\tilde{\rho}} \eta_j) +
\varepsilon^{ijk} (\bar{\lambda}_i \eta_k) \qquad
\delta \tilde{\pi}^j = - (\bar{\tilde{\rho}} \gamma_5 \eta_j) +
\varepsilon^{ijk} (\bar{\lambda}_i \gamma_5 \eta_k) \nonumber
\end{eqnarray}
the axial charges being:
\begin{center}
\begin{tabular}{|c|c|c|c|c|c|} \hline
field & $\Phi_\mu{}^i$ & $\chi^{ij}$, $\lambda_i$ & $z_i$,
$\tilde{z}_i$ & $\tilde{\rho}$ & others \\ \hline
axial charge & - 1 & + 1 & - 2 & - 3 & 0 \\ \hline
\end{tabular}
\end{center}

Now by using transformation properties of all fields we could decide
which fields will play the role of the Goldstone ones. For massive
spin-2 it will be $A_\mu$ and $\varphi$, for vector fields $B_\mu$,
$C_\mu{}^a$ and $(A_\mu + \gamma_5 B_\mu)^i$ --- scalar fields $\pi$,
$\pi^a$ and some combination of $z_i$ and $\tilde{z}_i$,
correspondingly. As for the gravitini fields, for the triplet
$\Psi_{\mu i}$ it could be some combination of $\lambda_i$ and
$\chi^{[ij]}$, while for the antitriplet $\Phi_\mu{}^i$ that of
$\lambda^i$ and $\chi_{[ij]}$. Then the most general mass terms
compatible with the $U(3)$ symmetry could be written as follows:
\begin{eqnarray}
\frac{1}{m} {\cal L}_1 &=& \sqrt{2} [ h^{\mu\nu} \partial_\mu A_\nu -
h (\partial A) ] - \sqrt{3} A^\mu \partial_\mu \varphi - B^\mu
\partial_\mu \pi + C_\mu{}^a \partial_\mu \pi^a - A_\mu{}^i \partial_
\mu \varphi_1{}^i - B_\mu{}^i \partial_\mu \pi_1{}^i - \nonumber \\
 && - \bar{\Phi}_\mu{}^i \sigma^{\mu\nu} \Psi_{\nu i} + i \kappa_1
 \bar{\Psi}_\mu{}^i \gamma^\mu \lambda_i + i \kappa_2
 \bar{\Psi}_\mu{}^i \gamma^\mu \varepsilon_{ijk} \chi^{jk} + i
 \kappa_3 \bar{\Phi}_{\mu i} \gamma^\mu \lambda^i + i \kappa_4
 \bar{\Phi}_{\mu i} \gamma^\mu \varepsilon^{ijk} \chi_{jk} + \nonumber
 \\
 && + \kappa_5 \bar{\chi}^{ij} \chi_{ij} + \kappa_6 \bar{\chi}^{ij}
 \chi_{ji} + \kappa_7 \varepsilon_{ijk} \bar{\chi}^{ij} \lambda^k +
 \kappa_8 \varepsilon^{ijk} \bar{\chi}_{ij} \lambda_k + \kappa_9 \bar{
 \lambda}^i \lambda_i + \kappa_{10} \bar{\rho} \tilde{\rho} \\
\frac{1}{m^2} {\cal L}_2 &=& - \frac{1}{2} ( h^{\mu\nu} h_{\mu\nu} -
h^2) - \sqrt{\frac{3}{2}} h \varphi + \varphi^2 + \frac{1}{2} B_\mu{}^
2 + \frac{1}{2} (C_\mu{}^a)^2 + \nonumber \\
 && + \frac{1}{2} (A_\mu{}^i)^2 + \frac{1}{2} (B_\mu{}^i)^2 -
 \frac{1}{2} (\varphi^a)^2 - \frac{1}{2} (\varphi_2{}^i)^2 -
 \frac{1}{2} (\pi_2{}^i)^2
\end{eqnarray}
here
$\varphi_1{}^i = cos\alpha \varphi^i + sin\alpha \tilde{\varphi}^i$,
$\varphi_2{}^i = - sin\alpha \varphi^i + cos\alpha \tilde{\varphi}^i$
and analogously for $\pi_1{}^i$ and $\pi_2{}^i$.

As usual one has to add also additional terms to the fermionic
transformation laws (see later). The requirement that the whole
massive theory be invariant under the $N=3$ global
supertransformations completely fixes the mixing angles as well as all
the unknown coefficients:
$$
sin\theta = \frac{\sqrt{3}}{2} \qquad cos\theta = \frac{1}{2} \qquad
sin\alpha = cos\alpha = - \frac{1}{\sqrt{2}}
$$
$$
\kappa_1 = \kappa_2 = \kappa_3 = \kappa_4 = \frac{1}{\sqrt{2}} \qquad
\kappa_6 = - \kappa_7 = - \kappa_8 = \kappa_{10} = 1 \qquad \kappa_5 =
\kappa_9 = 0
$$
The structure of the gravitini mass terms obtained corresponds to the
invariance under the two triplets of local gauge transformations:
$$
\delta \Psi_{\mu i} = \partial_\mu \eta_i \qquad \delta \Phi_\mu{}^i =
\frac{im}{2} \gamma_\mu \eta_i \qquad \delta \chi^{ij} =
\frac{m}{\sqrt{2}} \varepsilon^{ijk} \eta_k \qquad \delta \lambda_i =
\frac{m}{\sqrt{2}} \eta_i
$$
$$
\delta \Psi_{\mu i} = \frac{im}{2} \gamma_\mu \xi^i \qquad \delta
\Phi_\mu{}^i = \partial_\mu \xi^i \qquad \delta \chi_{ij} =
\frac{m}{\sqrt{2}} \varepsilon_{ijk} \xi^k \qquad \delta \lambda^i =
\frac{m}{\sqrt{2}} \xi^i
$$
Using these symmetries we can by making fields dependent local
transformations to bring the fermionic supertransformation laws to
most simple form:
\begin{eqnarray}
\delta \Psi_{\mu i} &=& - \sigma^{\alpha\beta} \partial_\alpha
h_{\beta\mu} \eta_i + \frac{i}{4} \varepsilon_{ijk} (\sigma(A -
\gamma_5 B))^j \gamma_\mu \eta^k + \nonumber \\
 && + \frac{1}{\sqrt{6}} \gamma_5 D_\mu
\pi \eta_i - \gamma_5 D_\mu \pi^a (\lambda^a)_i{}^j \eta_j -
\frac{m}{\sqrt{2}} \gamma_\mu \hat{A} \eta_i \nonumber \\
\delta \chi_{ij} &=& - \frac{1}{2\sqrt{2}} (\sigma(A + \gamma_5 B))_j
\eta_i  - \frac{i}{\sqrt{2}}
\varepsilon_{ikl} \gamma^\mu (\partial_\mu \varphi^a + \gamma_5 D_\mu
\pi^a) (\lambda^a)_k{}^j \eta_l + \nonumber \\
 && - \frac{i}{\sqrt{3}} \varepsilon_{ijk} \gamma^\mu (\partial_\mu
\varphi + \gamma_5 D_\mu \pi) \eta^k + i m \hat{A} \varepsilon^{ijk}
\eta_k + \frac{m}{2} (z_i -  \tilde{z}_i) \eta_j \nonumber \\
\delta \chi^{ij} &=& - \frac{1}{2\sqrt{2}} \varepsilon^{jkl} \gamma_5
(\sigma C)_k{}^i \eta_l - \frac{1}{4\sqrt{3}} (\sigma(\sqrt{3} A -
\gamma_5 B)) \varepsilon^{ijk} \eta_k - i \gamma^\mu D_\mu z_i \eta_j
- \nonumber \\
 && - \frac{m}{2\sqrt{3}} \varepsilon^{ijk} \varphi \eta_k +
 \frac{m}{\sqrt{2}} \varepsilon^{ikl} \varphi^a (\lambda^a)_k{}^j
 \eta_l \\
\delta \lambda_i &=& - \frac{1}{4} (\sigma(A + \sqrt{3} \gamma_5 B))
\eta_i - i \gamma^\mu D_\mu \tilde{z}_j \varepsilon^{ijk} \eta_k -
\frac{m}{2\sqrt{3}} \varphi \eta_i + \frac{m}{\sqrt{2}} \varphi^a
(\lambda^a)_i{}^j \eta_j \nonumber \\
\delta \lambda^i &=& - \frac{i}{\sqrt{3}} \gamma^\mu (\partial_\mu
\varphi - \gamma_5 D_\mu \pi) \eta_i - \frac{i}{\sqrt{2}} \gamma^\mu
(\partial_\mu \varphi^a - \gamma_5 D_\mu \pi^a) (\lambda^a)_i{}^j
\eta_j + \nonumber \\
 && + i m \hat{A} \eta_i - \frac{m}{2} \varepsilon^{ijk} (z_j -
\tilde{z}_j) \eta_k \nonumber \\
\delta \rho &=& - \frac{1}{2\sqrt{2}} (\sigma(A - \gamma_5 B))^i
\eta_i - \frac{m}{2} (z^{*i} - \tilde{z}^{*i}) \eta_i \qquad \delta
\tilde{\rho} = - i \gamma^\mu D_\mu \tilde{z}^{*i} \eta_i \nonumber
\end{eqnarray}
where $D_\mu \pi = \partial_\mu \pi - m B_\mu$ and so on.

As we have already mentioned at the beginning of this section the
massless limit of such model gives exactly the set of fields
corresponding to $N=6$ supergravity. But usual $N=6$ supergravity is
invariant under the real $SO(6)$ subgroup of the whole $SU(6)$ $R$
symmetry group, in this vector fields are transformed as 15-plet and a
singlet under $SO(6)$. The supertransformations given above clearly
shows that it is necessary to have dual mixing of this singlet vector
with one of the fifteen vector fields. Now we will show that this
corresponds to dual version of $N=6$ supergravity invariant under
$USp(6)$ group with the vector fields transforming as $14 + 1 + 1$.
First of all we combine fermions into the representations of $USp(6)$:
$\Psi_{\mu \hat{i}} = (\Psi_{\mu i}, \Phi_\mu{}^i)$,
$\lambda_{\hat{i}} = (\lambda_i,\lambda^i)$,
$\chi_{[\hat{i}\hat{j}\hat{k}]} =
(\rho,\chi_{ij},\chi^{ij},\tilde{\rho})$, where
$\hat{i} = 1,2,3,4,5,6$. Then we introduce 14-plet of vector fields
$C_\mu{}^{\hat{a}} = ((A_\mu + \gamma_5 B_\mu)^i, C_\mu{}^a)$, where
$\hat{a} = 1.2...14$. Analogously, we combine scalar fields into two
14-plets: $\varphi^{\hat{a}} = (z_2{}^i, \varphi^a)$ and
$\pi^{\hat{a}} = (z_1{}^i,\pi^a)$, where
$z_{1,2}{}^i = \frac{1}{\sqrt{2}}(z^i \pm \tilde{z}^i)$. At last we
introduce 14 skew symmetric $6 \times 6$ matrices:
$$
\Gamma^j = \left( \begin{array}{cc} - \varepsilon_{ijk} & 0 \\ 0 &
\varepsilon^{ijk} \end{array} \right) \qquad \Gamma^{3+j} = \left(
\begin{array}{cc} \varepsilon_{ijk} \gamma_5 & 0 \\ 0 & \varepsilon^{
ijk} \gamma_5 \end{array} \right) \qquad \Gamma^{6+a} = \left(
\begin{array}{cc} 0 & \gamma_5 (\lambda^a) \\ - \gamma_5 (\lambda^a) &
0 \end{array} \right)
$$
satisfying the relation
$\Omega \bar{\Gamma}^{\hat{a}} + \Gamma^{\hat{a}} \Omega = 0$, where
$\Omega$ is skew symmetric invariant tensor of $USp(6)$. Using this
variables one can uplift massless limit of supertransformations
obtained above up to full $N=6$ supersymmetry (omitting hats):
\begin{eqnarray}
\delta e_{\mu r} &=& i (\bar{\Psi}_\mu{}^i \gamma_r \eta_i) \nonumber
\\
\delta \Psi_{\mu i} &=& 2 D_\mu \eta_i - \frac{i}{4} (\sigma C)^a
\gamma_\mu (\bar{\Gamma}^a)^{ij} \eta_j - \frac{i}{4\sqrt{6}} (\sigma
(\sqrt{3} A - \gamma_5 B)) \gamma_\mu \Omega^{ij} \eta_j - \nonumber
\\
 && - \partial_\mu \pi^a \Omega_{ij} (\bar{\Gamma}^a)^{jk} \eta_k +
\frac{1}{\sqrt{6}} \gamma_5 \partial_\mu \pi \eta_i \nonumber \\
\delta A_\mu &=& \frac{1}{\sqrt{2}} (\bar{\Psi}_{\mu i} \Omega^{ij}
\eta_j) - \frac{i\sqrt{6}}{4} (\bar{\chi}^{ijk} \gamma_\mu \Omega_{ij}
\eta_k) + \frac{i}{2} (\bar{\lambda}^i \gamma_\mu \eta_i) \nonumber \\
\delta B_\mu &=& \frac{1}{\sqrt{6}} (\bar{\Psi}_{\mu i} \gamma_5
\Omega^{ij} \eta_j) + \frac{i}{2\sqrt{2}} (\bar{\chi}^{ijk} \gamma_\mu
\gamma_5 \Omega_{ij} \eta_k) + \frac{i\sqrt{3}}{2} (\bar{\lambda}^i
\gamma_\mu \gamma_5 \eta_i) \nonumber \\
\delta C_\mu{}^a &=& (\bar{\Psi}_{\mu i} (\bar{\Gamma}^a)^{ij} \eta_j)
- \frac{i\sqrt{3}}{2} (\bar{\chi}^{ijk} \gamma_\mu (\Gamma^a)_{ij}
\eta_k) \\
\delta \chi_{ijk} &=& \frac{\sqrt{3}}{4} (\sigma C)^a (\Gamma^a)_{[ij}
\eta_{k]} + \frac{1}{4\sqrt{2}} (\sigma (\sqrt{3} A - \gamma_5 B))
\Omega_{[ij} \eta_{k]} - \nonumber \\
 && - \frac{i}{\sqrt{2}} \gamma^\mu \partial_\mu (\varphi + \gamma_5
 \pi) \Omega^{[ij} \Omega^{k]l} \eta_l + \frac{i\sqrt{3}}{2}
 \gamma^\mu \partial_\mu (\pi^a + \gamma_5
\varphi^a) (\bar{\Gamma}^a)^{[ij} \Omega^{k]l} \eta_l \nonumber \\
\delta \lambda_i &=& - \frac{1}{4} (\sigma (A + \sqrt{3} \gamma_5 B))
\eta_i - \frac{i}{\sqrt{3}} \gamma^\mu \partial_\mu (\varphi -
\gamma_5 \pi) \Omega^{ij} \eta_j + \nonumber \\
 && + \frac{i}{\sqrt{2}} \gamma^\mu \partial_\mu (\pi^a - \gamma_5
 \varphi^a) (\bar{\Gamma}^a)^{ij} \eta_j \nonumber \\
\delta \varphi &=& \frac{1}{\sqrt{3}} (\bar{\lambda}_i \Omega^{ij}
\eta_j) + \frac{1}{\sqrt{2}} (\bar{\chi}_{ijk} \Omega^{ij} \Omega^{kl}
\eta_l) \nonumber \\
\delta \pi &=& - \frac{1}{\sqrt{3}} (\bar{\lambda}_i
\gamma_5 \Omega^{ij} \eta_j) + \frac{1}{\sqrt{2}} (\bar{\chi}_{ijk}
\gamma_5 \Omega^{ij} \Omega^{kl} \eta_l) \nonumber \\
\delta \varphi^a &=& \frac{1}{\sqrt{2}} (\bar{\lambda}_i \gamma_5
(\bar{\Gamma}^a)^{ij} \eta_j) - \frac{\sqrt{3}}{2} (\bar{\chi}_{ijk}
\gamma_5 (\bar{\Gamma}^a)^{ij} \Omega^{kl} \eta_l) \nonumber \\
\delta \pi^a &=& - \frac{1}{\sqrt{2}} (\bar{\lambda}_i
(\bar{\Gamma}^a)^{ij} \eta_j) - \frac{\sqrt{3}}{2} (\bar{\chi}_{ijk}
(\bar{\Gamma}^a)^{ij} \Omega^{kl} \eta_l) \nonumber
\end{eqnarray}

In principle by straightforward calculations one can construct full
nonlinear interacting model corresponding to such dual version of
$N=6$ supergravity. In this, the scalar fields $\pi^a$ will be related
with the global translations $\pi^a \rightarrow \pi + \Lambda^a$,
while the scalar fields $\varphi^a$ will realize the nonlinear
$\sigma$-model $SU(6)^*/USp(6)$. Ones again, these properties together
with the $USp(6)$ symmetry show the connection of such theory with the
$N=6$ $D=5$ supergravity.

\section{$N=4$}

For $N=4$ massive spin-2 supermultiplet contains the following fields:
$(2,8 \times 3/2,  27 \times 1, 48 \times 1/2, 42 \times 0)$. In the
massless limit one has additionally $(1, 8 \times 1/2, 28 \times 0)$
which together gives exactly the $N=8$ supergravity multiplet
$(2, 8 \times 3/2, 28 \times 1, 56 \times 1/2, 70 \times 0)$. In terms
of massless $N=4$ supermultiplets it corresponds to $N=4$ supergravity
multiplet $(2, 4 \times 3/2, 6 \times 1, 4 \times 1/2, 2 \times 0)$,
four spin-3/2 multiplets
$ 4 \times (3/2, 4 \times 1, 7 \times 1/2, 8 \times 0)$ and six vector
supermultiplets $6 \times (1, 4 \times 1/2, 6 \times 0)$.

As is well known usual $N=4$ supergravity as a maximum could have
$SU(4)$ symmetry. But as we have shown long time ago \cite{Z87a} there
exist dual version of the system $N=4$ supergravity plus six vector
supermultiplets with the complete $U(4)$ symmetry. It has been shown
that such theory admits spontaneous supersymmetry breaking with four
arbitrary scales and without a cosmological term, including all
partial super-Higgs effects $N=4 \rightarrow N=3$,
$N=4 \rightarrow N=2$ and $N=4 \rightarrow N=1$. Moreover, this theory
could be coupled to arbitrary number of vector supermultiplets
\cite{TZ96b}. So our first building block will be this dual version of
$N=4$ supergravity multiplet mixed with six vector supermultiplet (see
Appendix for notations):
\begin{eqnarray}
\delta h_{\mu\nu} &=& i (\bar{\Psi}_{(\mu}{}^i \gamma_{\nu)} \eta_i)
\nonumber \\
\delta \Psi_{\mu i} &=& - \sigma^{\alpha\beta} \partial_\alpha
h_{\beta\mu} \eta_i - \frac{i}{4\sqrt{2}} (\sigma V^*)^a \gamma_\mu
(\bar{\tau}^a)^{ij} \eta_j \nonumber \\
\delta A_\mu{}^a &=& \frac{1}{\sqrt{2}} (\bar{\Psi}_\mu{}^i
(\bar{\tau}^a)^{ij} \eta_j) - \frac{i}{\sqrt{2}} (\bar{\chi}^{ai}
\gamma_\mu \eta_i) + \frac{i}{2} (\bar{\lambda}_i \gamma_\mu
(\bar{\tau}^a)^{ij} \eta_j) \nonumber \\
\delta B_\mu{}^a &=& \frac{1}{\sqrt{2}} (\bar{\Psi}_\mu{}^i \gamma_5
(\bar{\tau}^a)^{ij} \eta_j) - \frac{i}{\sqrt{2}} (\bar{\chi}^{ai}
\gamma_\mu \gamma_5 \eta_i) - \frac{i}{2} (\bar{\lambda}_i \gamma_\mu
\gamma_5 (\bar{\tau}^a)^{ij} \eta_j) \\
\delta \chi_i{}^a &=& \frac{1}{2\sqrt{2}} (\sigma V)^a \eta_i - i
\gamma^\mu \partial_\mu z^{ab} (\bar{\tau}^b)^{ij} \eta_j \nonumber \\
\delta \lambda^i &=& - \frac{1}{4} (\sigma V^*)^a (\bar{\tau}^a)^{ij}
\eta_j - i \gamma^\mu \partial_\mu z \eta_i \nonumber \\
\delta \varphi &=& (\bar{\lambda}^i \eta_i) \qquad \delta \pi =
(\bar{\lambda}^i \gamma_5 \eta_i) \qquad \delta z^{ab} =
(\bar{\chi}_i{}^a (\bar{\tau}^b)^{ij} \eta_j) \nonumber
\end{eqnarray}
where $V_{\mu\nu}{}^a = A_{\mu\nu}{}^a + \gamma_5 B_{\mu\nu}{}^a$.

The remaining fields are just four spin-3/2 supermultiplets. By the
same reasoning as in the $N=2$ and $N=3$ cases we choose gravitini to
be transformed as $\Phi_\mu{}^i$ under $SU(4)$. In this, vector fields
will be transformed as 15-plet and a singlet. There are no any other
vector supermultiplets left, so we can not make any dual mixing with
this singlet vector field, but there exist two possibilities when this
singlet field is a vector or axial vector one. Having in mind the role
of this field as the (only) candidate for the role of the Goldstone
field for massive spin-2, we choose it to be a vector. Then the
supertransformations for all fields of these four spin-3/2
supermultiplets look as follows:
\begin{eqnarray}
\delta \Phi_\mu{}^i &=& - \frac{i}{8} (\sigma C)^{ab} \gamma_\mu
(\Sigma^{ab})_i{}^j \eta_j - \frac{i}{4\sqrt{2}} (\sigma A) \gamma_\mu
\eta_i \nonumber \\
\delta C_\mu{}^{ab} &=& \frac{1}{2} (\bar{\Phi}_\mu{}^i
(\Sigma^{ab})_i{}^j \eta_j) + \frac{i}{4} (\bar{\Omega}^{ic}
\gamma_\mu (\Sigma^{ab})_j{}^i (\bar{\tau}^c)^{jk} \eta_k) \nonumber
\\
\delta A_\mu &=& \frac{1}{\sqrt{2}} (\bar{\Phi}_\mu{}^i \eta_i) -
\frac{i}{2\sqrt{2}} (\bar{\Omega}_i{}^a \gamma_\mu (\bar{\tau}^a)^{ij}
\eta_j) \nonumber \\
\delta \Omega^{ia} &=& - \frac{1}{8} (\sigma C)^{bc}
(\Sigma^{bc})_j{}^i (\bar{\tau}^a)^{jk} \eta_k + \frac{1}{4\sqrt{2}}
(\sigma A) (\bar{\tau}^a)^{ij} \eta_j - \nonumber \\
 && - i \gamma^\mu \partial_\mu [ \frac{1}{2\sqrt{2}} (\tau^b)_{ij}
 (\bar{\tau}^a)^{jk} z^b + \frac{1}{4\sqrt{3}} (\Gamma^{bcd})_{ij}
 (\bar{\tau}^a)^{jk} z^{bcd} ] \eta_k \\
\delta \rho_i &=& - i \gamma^\mu \partial_\mu [ \frac{1}{2}
(\bar{\tau}^a)^{ij} z^{*a} + \frac{1}{2\sqrt{6}}
(\bar{\Gamma}^{abc})^{ij} z^{abc} ] \eta_j \nonumber \\
\delta \varphi^a &=& \frac{1}{2\sqrt{2}} (\bar{\Omega}^{ib}
(\tau^a)_{ij} (\bar{\tau}^b)^{jk} \eta_k) + \frac{1}{2} (\bar{\rho}_i
(\bar{\tau}^a)^{ij} \eta_j) \nonumber \\
\delta \pi^a &=& \frac{1}{2\sqrt{2}} (\bar{\Omega}^{ib} \gamma_5
(\tau^a)_{ij} (\bar{\tau}^b)^{jk} \eta_k) - \frac{1}{2} (\bar{\rho}_i
\gamma_5 (\bar{\tau}^a)^{ij} \eta_j) \nonumber \\
\delta z^{abc} &=& \frac{1}{4\sqrt{3}} (\bar{\Omega}^{id}
(\Gamma^{abc})_{ij} (\bar{\tau}^a)^{jk} \eta_k) + \frac{1}{2\sqrt{6}}
(\bar{\rho}_i (\bar{\Gamma}^{abc})^{ij} \eta_j) \nonumber
\end{eqnarray}

In this, all nonzero axial charges will be:
\begin{center}
\begin{tabular}{|c|c|c|c|c|c|c|} \hline
field & $\eta_i$, $\Psi_{\mu i}$, $\Omega^{ia}$ & $\Phi_\mu{}^i$,
$\chi_i{}^a$ & $(A_\mu + \gamma_5 B_\mu)^a$, $z^a$ & $\lambda^i$ &
$\rho_i$ & $z$ \\ \hline
axial charge & + 1 & - 1 & - 2 & + 3 & - 3 & - 4 \\ \hline
\end{tabular}
\end{center}

Accordingly, the role of Goldstone fields will be played by $A_\mu$
and $Tr(\varphi^{ab})$ for massive spin-2, by $z^a$ and $\pi^{ab}$ for
$(A_\mu + \gamma_5 B_\mu)^a$ and $C_\mu{}^{ab}$, and by
$(\tau^a)_{ij} \Omega^{ja}$ and $(\bar{\tau}^a)^{ij} \chi_j{}^a$ for
$\Psi_{\mu i}$ and $\Phi_\mu{}^i$, correspondingly. That leads to the
following most general form of the mass terms in the Lagrangian:
\begin{eqnarray}
\frac{1}{m} {\cal L}_1 &=& \sqrt{2} [ h^{\mu\nu} \partial_\mu A_\nu -
h (\partial A)] - \sqrt{3} A^\mu \partial_\mu \varphi_0 + C_\mu{}^{ab}
\partial_\mu \pi^{ab} - A_\mu{}^a \partial_\mu \varphi^a - B_\mu{}^a
\partial_\mu \pi^a \nonumber \\
 && - \bar{\Phi}_\mu{}^i \sigma^{\mu\nu} \Psi_{\nu i} + i \kappa_1
 \bar{\Psi}_\mu{}^i \gamma^\mu (\tau^a)_{ij} \Omega^{ja} + i \kappa_2
 \bar{\Phi}_{\mu i} \gamma^\mu (\bar{\tau}^a)^{ij} \chi_j{}^a +
 \nonumber \\
 && + \kappa_3 \bar{\Omega}^{ia} \chi_i{}^a + \kappa_4
 \bar{\Omega}^{ia} (\tau^a)_{ij} (\bar{\tau}^b)^{jk} \chi_k{}^b +
 \kappa_5 \bar{\lambda}^i \rho_i \\
\frac{1}{m^2} {\cal L}_2 &=& - \frac{1}{2} (h^{\mu\nu} h_{\mu\nu} -
h^2) - \sqrt{\frac{3}{2}} h \varphi_0 + \varphi_0{}^2 + \frac{1}{2}
(C_\mu{}^{ab})^2 + \frac{1}{2} (A_\mu{}^a)^2 + \frac{1}{2}
(B_\mu{}^a)^2 - \nonumber \\
 && - \frac{1}{2} (\hat{\varphi}^{ab})^2 - \frac{1}{2} (z^{abc})^2 -
 \frac{1}{2} z^* z
\end{eqnarray}
where $\varphi_0 = Tr(\varphi^{ab})$,
$\hat{\varphi}^{ab} = \varphi^{ab} - 1/6 \delta^{ab} \varphi_0$. By
adding appropriate terms to the fermionic transformation laws (see
later) global $N=4$ supersymmetry could be restored, which requires:
$$
\kappa_1 = \frac{1}{2} \qquad \kappa_2 = \kappa_4 = - \frac{1}{2}
\qquad \kappa_3 = \kappa_5 = - 1
$$
and fixes all the coefficients in the supertransformations. The
structure of the gravitini mass terms corresponds to the invariance
under the two quartets of local gauge transformations:
$$
\delta \Psi_{\mu i} = \partial_\mu \eta_i \qquad \delta \Phi_\mu{}^i =
\frac{im}{2} \gamma_\mu \eta_i \qquad \delta \Omega^{ia} = -
\frac{m}{2} (\bar{\tau}^a)^{ij} \eta_j
$$
$$
\delta \Psi_{\mu i} = \frac{im}{2} \gamma_\mu \xi^i \qquad \delta
\Psi_\mu{}^i = \partial_\mu \xi^i \qquad \delta \chi_i{}^a =
\frac{m}{2} (\tau^a)_{ij} \xi^j
$$
As usual, we use this freedom to bring the supertransformation laws to
most simple form:
\begin{eqnarray}
\delta \Psi_{\mu i} &=& - \sigma^{\alpha\beta} \partial_\alpha
h_{\beta\mu} \eta_i - \frac{i}{4\sqrt{2}} (\sigma V^*)^a \gamma_\mu
(\bar{\tau}^a)^{ij} \eta_j - \frac{1}{2} D_\mu \pi^{ab}
(\Sigma^{ab})_i{}^j \eta_j - \frac{m}{\sqrt{2}} \gamma_\mu \hat{A}
\eta_i \nonumber \\
\delta \Phi_\mu{}^i &=& - \frac{i}{8} (\sigma C)^{ab} \gamma_\mu
(\Sigma^{ab})_i{}^j \eta_j - \frac{i}{4\sqrt{2}} (\sigma A) \gamma_\mu
\eta_i + \frac{1}{\sqrt{2}} D_\mu z^a (\bar{\tau}^a)^{ij} \eta_j +
\nonumber \\
 && + im h_{\mu\nu} \gamma^\nu \eta_i - \frac{im\sqrt{6}}{4} \gamma_
 \mu \varphi_0 \eta_i \nonumber \\
\delta \chi_i{}^a &=& \frac{1}{2\sqrt{2}} (\sigma V)^a \eta_i - i
\gamma^\mu [ \partial_\mu \varphi^{ab} + D_\mu \pi^{ab} ] (\bar{\tau}^
b)^{ij} \eta_j + \nonumber \\
 && + \frac{im}{\sqrt{2}} \hat{A} (\bar{\tau}^a)^{ij}
\eta_j - \frac{m}{4\sqrt{3}} z^{bcd} (\Gamma^{bcd})_{ij}
(\bar{\tau}^a)^{jk} \eta_k \nonumber \\
\delta \Omega^{ia} &=& - \frac{1}{8} (\sigma C)^{bc}
(\Sigma^{bc})_j{}^i (\bar{\tau}^a)^{jk} \eta_k + \frac{1}{4\sqrt{2}}
(\sigma A) (\bar{\tau}^a)^{ij} \eta_j -  \\
 && - i \gamma^\mu [ \frac{1}{2\sqrt{2}} (\tau^b)_{ij}
 (\bar{\tau}^a)^{jk} D_\mu z^b + \frac{1}{4\sqrt{3}}
 (\Gamma^{bcd})_{ij} (\bar{\tau}^a)^{jk} \partial_\mu z^{bcd} ] \eta_k
 + \nonumber \\
 && + \frac{m}{2\sqrt{6}} \varphi_0 (\bar{\tau}^a)^{ij} \eta_j - m
 \hat{\varphi}^{ab} (\bar{\tau}^b)^{ij} \eta_j \nonumber \\
\delta \lambda^i &=& - \frac{1}{4} (\sigma V^*)^a (\bar{\tau}^a)^{ij}
\eta_j - i \gamma^\mu \partial_\mu z \eta_i - \frac{m}{2\sqrt{6}}
z^{abc} (\bar{\Gamma}^{abc})^{ij} \eta_j \nonumber \\
\delta \rho_i &=& - i \gamma^\mu [ \frac{1}{2}
(\bar{\tau}^a)^{ij} D_\mu z^{*a} + \frac{1}{2\sqrt{6}}
(\bar{\Gamma}^{abc})^{ij} \partial_\mu z^{abc} ] \eta_j - m z \eta_i
\nonumber
\end{eqnarray}

As we have already noted the set of fields corresponds to $N=8$
supergravity multiplet. Now we consider massless limit of the
supertransformations obtained and try to uplift them upto $N=8$
supertransformations to see which dual version of $N=8$ supergravity
corresponds to our case. First af all we combine eight gravitini into
octet $\Psi_{\mu \hat{i}}$, $\hat{i} = 1,2,...8$, vector fields
$A_\mu{}^a$, $B_\mu{}^a$ and $C_\mu{}^{ab}$ into 27-plet $C_\mu{}^A$,
$A=1,2...27$, leaving $A_\mu$ as a singlet, all spinor fields --- into
completely skew symmetric third rank tensor
$\chi_{[\hat{i}\hat{j}\hat{k}]}$. As for the scalar fields they could
be organized into the singlet $\varphi_0$, 27-plet $z^A$ and
completely skew symmetric fourth rank $\Omega$-traceless tensor
$\Phi^{[\hat{i}\hat{j}\hat{k}\hat{l}]}$, where
$\Omega^{[\hat{i}\hat{j}]}$ --- skew symmetric invariant tensor of
$USp(8)$. At last we introduce 27 skew symmetric matrices
$(\Gamma^A)_{[\hat{i}\hat{j}]}$:
$$
\Gamma^a = \left( \begin{array}{cc} \frac{1}{\sqrt{2}} \tau^a & 0 \\ 0
& - \frac{1}{\sqrt{2}} \bar{\tau}^a \end{array} \right) \quad
\Gamma^{6+a} = \left( \begin{array}{cc} - \frac{1}{\sqrt{2}} \gamma_5
\tau^a & 0 \\ 0 & - \frac{1}{\sqrt{2}} \gamma_5 \bar{\tau}^a
\end{array} \right) \quad \Gamma^{12+(ab)} = \left( \begin{array}{cc}
0 & \frac{1}{2} \Sigma^{ab} \\ - \frac{1}{2} \Sigma^{ab} & 0
\end{array} \right)
$$
Using these notations one can obtain the following $USp(8)$-invariant
supertransformations (omitting hats):
\begin{eqnarray}
\delta e_{\mu r} &=& i (\bar{\Psi}_\mu{}^i \gamma_r \eta_i) \nonumber
\\
\delta \Psi_{\mu i} &=& 2 D_\mu \eta_i - \frac{i}{4} (\sigma C)^A
\gamma_\mu (\bar{\Gamma}^A)^{ij} \eta_j - \frac{i}{4\sqrt{2}} (\sigma
A) \gamma_\mu \Omega^{ij} \eta_j + \partial_\mu z^A \Omega_{ij}
(\bar{\Gamma}^A)^{jk} \eta_k \nonumber \\
\delta C_\mu{}^A &=& (\bar{\Psi}_{\mu i} (\bar{\Gamma}^A)^{ij} \eta_j)
+ \frac{i\sqrt{3}}{2} (\bar{\chi}^{ijk} \gamma_\mu (\Gamma^A)_{ij}
\eta_k) \nonumber \\
\delta A_\mu &=& \frac{1}{\sqrt{2}} (\bar{\Psi}_{\mu i} \Omega^{ij}
\eta_j) + \frac{i\sqrt{6}}{4} (\bar{\chi}^{ijk} \gamma_\mu \Omega_{ij}
\eta_k) \\
\delta \chi_{ijk} &=& - \frac{\sqrt{3}}{4} (\sigma C)^A
(\Gamma^A)_{[ij} \eta_{k]}) - \frac{\sqrt{6}}{8} (\sigma A)
\Omega_{[ij} \eta_{k]} - \nonumber \\
 && - i \gamma^\mu \partial_\mu [ \varphi_0 \Omega^{[ij} \Omega^{k]l}
 + z^A (\bar{\Gamma}^A)^{[ij} \Omega^{k]l} + \Phi^{ijkl} ] \eta_l
 \nonumber \\
\delta \varphi_0 &=& (\bar{\chi}_{ijk} \Omega^{ij} \Omega^{kl} \eta_l)
\qquad \delta z^A = (\bar{\chi}_{ijk} (\bar{\Gamma}^A)^{ij}
\Omega^{kl} \eta_l) \nonumber
\end{eqnarray}

The full interacting version of $N=8$ supergravity with all required
properties ($USp(8)$-invariance, vector fields in 27-plet and a
singlet, 27-plet of scalar fields $z^A$ entering through the
derivatives $\partial_\mu z^A$ only and 42-plet $\Phi^{ijkl}$,
realizing non-linear $\sigma$-model $E_{6,6}/USp(8)$) is already known
\cite{SN82,GRW86,ADFL02b}. As in all previous cases it arises from the
dimensional reduction of five-dimensional supergravity.

\section*{Conclusion}

Let us summarize briefly the results of our work. First of all we have
seen that the framework based on gauge invariant description of
massive gauge particles allows one easily construct massive
supermultiplets out of the massless ones. We have seen also that in
all cases the possibility to make dual transformations on vector
fields plays a very important role. One of the properties of all the
models constructed is that the Lagrangian and supertransformations
turn out to be invariant under the whole $U(N)$ $R$ symmetry group of
$N$ extended superalgebra. As a consequence, the mass terms for the
gravitini appears to be the Dirac ones. It resembles very much the
situation in supersymmetric gauge theories. As is very well known, the
superpartners for the massless vector fields such as photon and gluons
are Majorana spinors, while superpartners for the massive one such as
$W$ and $Z$ bosons are the Dirac spinors. Analogously, we have shown
that the superpartners for massive graviton are Dirac gravitini.

For the massless limits of all four cases $N=k$, $k=1,2,3,4$, we have
managed to uplift supertransformations up to $N=2k$ supersymmetry. We
did not try to give an exhaustive classification of all possible
extended supergravity models which such massive supermultiplets could
in principle originate from. But we have shown that there exist at
least examples of such theories having desired properties. All this
examples are the ones that can be obtained by dimensional reduction
from five-dimensional supergravities.

\appendix

\section{Notations and some useful formulas}

In this Appendix we collect our notations, conventions and some
useful formulas.

For the flat Minkowski space we use the metric
$g_{\mu\nu} = diag(+,-,-,-)$. Throughout the paper we use Majorana
representation of Dirac gamma matrices $\gamma_\mu$ in which all of
them are imaginary, the charge conjugation matrix being just
$\gamma_0$. In this, combinations $\gamma_0 \gamma_\mu$ and
$\gamma_0 \sigma_{\mu\nu}$ are symmetric in their spinor indices,
while $\gamma_0$, $\gamma_0 \gamma_5$ and
$\gamma_0 \gamma_5 \gamma_\mu$ --- skew symmetric. In this
representation Majorana spinors are real, so the $U(N)$ $R$ symmetry
group is a so called Majorana $U(N)$, where imaginary unit $i$ is
replaced by $\gamma_5$. In particular, the simplest $U(1)$ symmetry is
nothing else but axial transformations acting on spinor fields as
$$
\Psi \rightarrow e^{q \gamma_5 \Lambda} \Psi
$$
where $q$ --- axial charge of this field.

\subsection{Massive spin-2 particle}

In order to have gauge invariant description of massive spin-2
particle with a non-singular massless limit one has to introduce three
fields: symmetric tensor $h_{(\mu\nu)}$, vector $A_\mu$ and scalar
$\varphi$. We start with the sum of standard free massless
Lagrangians:
$$
{\cal L}_0 = \frac{1}{2} \partial^\alpha h^{\mu\nu} \partial_\alpha
h_{\mu\nu} - (\partial h)^\mu (\partial h)_\mu + (\partial h)^\mu
\partial_\mu h - \frac{1}{2} \partial^\mu h \partial_\mu h  -
\frac{1}{4} (A_{\mu\nu})^2 + \frac{1}{2} \partial^\mu \varphi
\partial_\mu \varphi
$$
where $A_{\mu\nu} = \partial_\mu A_\nu - \partial_\nu A_\mu$. Now by
adding terms with one and no derivatives to the Lagrangian and
appropriate corrections to the gauge transformations one can easily
obtain gauge invariant formulation:
$$
{\cal L}_1 = m \sqrt{2} [ h^{\mu\nu} \partial_\mu A_\nu - h
(\partial A) ] - m \sqrt{3} A^\mu \partial_\mu \varphi - \frac{m^2}{2}
(h^{\mu\nu} h_{\mu\nu} - h^2) - m^2 \sqrt{\frac{3}{2}} h \varphi + m^2
\varphi^2
$$
in this, the total Lagrangian is invariant under the following gauge
transformations:
$$
\delta h_{\mu\nu} = \partial_\mu \xi_\nu + \partial_\nu \xi_\mu +
\frac{m}{\sqrt{2}} g_{\mu\nu} \Lambda \qquad \delta A_\mu = \partial_
\mu \Lambda + m \sqrt{2} \xi_\mu \qquad \delta \varphi = m \sqrt{3}
\Lambda
$$

\subsection{Massless supermultiplet $(2,3/2)$}

As is well known, in supergravity theories (and in general in all
gravity theories with spinor fields) one usually uses so called tetrad
formulation of gravity in terms of tetrad $e_\mu{}^a$ and Lorentz
connection $\omega_\mu{}^{ab}$. But at the level of free models
considered here it turns out to be convenient to use simply symmetric
tensor $h_{\mu\nu}$. In this, the global supertransformations leaving
a sum of massless spin-2 and spin-3/2 Lagrangians invariant has the
form:
$$
\delta h_{\mu\nu} = i (\bar{\Psi}_{(\mu} \gamma_{\nu)} \eta) \qquad
\delta \Psi_\mu = - \sigma^{\alpha\beta} \partial_\alpha h_{\beta\mu}
\eta
$$
One can easily check that the commutator of two supertransformations
gives:
$$
[\delta_1,\delta_2] h_{\mu\nu} = - 2 i (\bar{\eta}_2 \gamma^\alpha
\eta_1) \partial_\alpha h_{\mu\nu} + \partial_\mu \xi_\nu +
\partial_\nu \xi_\mu
$$
where $\xi_\mu = 2 i (\bar{\eta}_2 \gamma^\alpha h_{\alpha\mu}
\eta_1)$.

\subsection{Dirac mass term for gravitini}

It is very well known that by using Goldstone spinor fields one can
construct gauge invariant formulation for one Majorana spin-3/2
particle (gravitino). Indeed, the Lagrangian:
$$
{\cal L} = \frac{i}{2} \varepsilon^{\mu\nu\alpha\beta} \bar{\Psi}_\mu
\gamma_5 \gamma_\nu \partial_\alpha \Psi_\beta + \frac{i}{2}
\bar{\chi} \gamma^\mu \partial_\mu \chi - \frac{m}{2} \bar{\Psi}_\mu
\sigma^{\mu\nu} \Psi_\nu + i m \sqrt{\frac{3}{2}} \bar{\Psi}_\mu
\gamma^\mu \chi + \bar{\chi} \chi
$$
is invariant under the gauge transformations:
$$
\delta \Psi_\mu = \partial_\mu \eta + \frac{im}{2} \gamma_\mu \eta
\qquad \delta \chi = m \sqrt{\frac{3}{2}} \eta
$$

But if we have two gravitini with equal masses there exist two
possibilities. The first one is just the sum of two copies of the
Lagrangian given above. Another one is a Dirac mass term
$m \bar{\Psi}_\mu \sigma^{\mu\nu} \Phi_\nu$, which could be
diagonalized into
$- \frac{m}{2} \bar{\Psi}_\mu \sigma^{\mu\nu} \Psi_\nu + \frac{m}{2}
\bar{\Phi}_\mu \sigma^{\mu\nu} \Phi_\nu$ (note the sign difference).
The Lagrangian for such case can also be constructed in a similar
manner:
$$
{\cal L} = {\cal L}_0 (\Psi_\mu, \Phi_\mu, \chi, \lambda) - m \bar{
\Psi}_\mu \sigma^{\mu\nu} \Phi_\nu + im \sqrt{\frac{3}{2}} (\bar{\Psi}
\gamma) \chi + im \sqrt{\frac{3}{2}} (\bar{\Phi} \gamma) \lambda - 2
\bar{\chi} \lambda
$$
which is invariant under the two gauge transformations:
$$
\delta \Psi_\mu = \partial_\mu \eta + \frac{im}{2} \gamma_\mu \xi
\qquad \delta \Phi_\mu = \partial_\mu \xi + \frac{im}{2} \gamma_\mu
\eta \qquad \delta \chi = m \sqrt{\frac{3}{2}} \eta \qquad \delta
\lambda = m \sqrt{\frac{3}{2}} \xi
$$
In all massive spin-2 supermultiplets we have pairs of gravitini with
opposite axial charges so the Dirac mass term turns out to be the only
possibility.

\subsection{$SO(6) \approx SU(4)$ matrices}

For $N=4$ supersymmetry we use six skew symmetric matrices
$(\tau^a)_{[ij]}$, $a=1,2,3,4,5,6$, $i,j=1,2,3,4$, satisfying the
relation:
$$
(\tau^a)_{ij} (\bar{\tau}^b)^{jk} + (\tau^b)_{ij} (\bar{\tau}^a)^{jk}
= - 2 \delta_i{}^k \delta^{ab}
$$
where
$(\bar{\tau}^a)^{ij} = \frac{1}{2} \varepsilon^{ijkl} (\tau^a)_{kl}$.
The commutator $\tau$-matrices defines 15 antihermitian matrices:
$$
(\Sigma^{[ab]})_i{}^k = \frac{1}{2} [ (\tau^a)_{ij}
(\bar{\tau}^b)^{jk} - (\tau^b)_{ij} (\bar{\tau}^a)^{jk} ]
$$
which play the role of $SO(6) \approx SU(4)$ generators. Besides, we
need 20 symmetric matrices:
$$
(\Gamma^{[abc]})_{(ij)} = (\tau^{[a})_{ik} (\bar{\tau}^b)^{kl}
(\tau^{c]})_{lj}
$$
which are self-dual in a sense that:
$$
\Gamma^{abc} = \frac{1}{6} \gamma_5 \varepsilon^{abcdef} \Gamma^{def}
$$
Let us give here some useful formulas with these matrices:
$$
(\bar{\tau}^a)^{ij} (\tau^b)_{jk} (\bar{\tau}^a)^{kl} = 4
(\bar{\tau}^b)^{il}
$$
$$
(\bar{\tau}^c)^{ij} (\Sigma^{ab})_j{}^k (\tau^c)_{kl} = 2
(\Sigma^{ab})_l{}^i
$$
$$
(\bar{\tau}^d)^{ij} (\Gamma^{abc})_{jk} (\bar{\tau}^d)^{kl} = 0
$$

\end{document}